\documentclass[11pt,aps,pra,tightenlines,nofootinbib,notitlepage,longbibliography]{revtex4-1}

\usepackage[T1]{fontenc}

\usepackage{amsmath}
\usepackage{amssymb}
\usepackage{bbm}
\usepackage{simplewick}
\usepackage{mathtools}
\usepackage[normalem]{ulem}
\usepackage{dsfont}
\usepackage{mathrsfs}
\usepackage[makeroom]{cancel}

\def\x{\boldsymbol{x}}
\def\y{\boldsymbol{y}}
\def\z{{\boldsymbol z}}

\def\o{{\text{o}}}

\def\0{\boldsymbol{0}}

\def\tchi{{\tilde\chi}}

\def\J{\boldsymbol{J}}

\def\q{\boldsymbol{q}}
\def\p{{\boldsymbol{p}}}

\DeclareMathOperator{\arctanh}{arctanh}

\def\E{\boldsymbol{E}}

\def\Bold{\boldsymbol{B}}

\def\T{\mathds{T}}
\def\LT{{ T}}

\def\ii{\mathrm{i}}

\def\d#1{d^3\mkern-1.5mu#1\,}

\def\dd#1{d^4\mkern-1.5mu#1\,}

\def\ddd#1{\frac{d^3\mkern-1.5mu#1}{(2\pi)^3}}
\def\dddd#1{\frac{d^4\mkern-1.5mu#1}{(2\pi)^4}}

\def\sig{{\boldsymbol\sigma}}

\newcommand{\rfock}[3]{
\ifthenelse{\equal{#2}{}}{|#1,\lfloor#3\rfloor\ra}{|#1,\lceil#2\rceil,\lfloor#3\rfloor\ra}
}
\newcommand{\lfock}[3]{
\ifthenelse{\equal{#2}{}}{\la#1,\lfloor#3\rfloor|}{\la#1,\lceil#2\rceil,\lfloor#3\rfloor|}
}

\def\eo{e_\o}
\def\mo{m_\o}

\def\vareps#1{\varepsilon_{\boldsymbol{#1}}}

\def\om#1{\omega_{\boldsymbol{#1}}}

\def\xxii{{\frac{1-\xi}{\xi}}}
\def\xxiidwa{{\frac{1-\xi}{2\xi}}}

\def\asym{\text{asym}}
\def\sym{\text{sym}}

\def\IN{{\rm int}}

\def\la{\langle}
\def\ra{\rangle}

\def\barelimT{T\to\infty(1-\ii0)} 
\def\limT{\lim_{T\to\infty(1-\ii0)}} 

\def\limlzero{\lim_{\substack{\lambda\to0}}}
\def\intTT{\int^T_{-T}}
\def\intT{\int_T}

\def\TT#1{\left.#1\mkern1mu\right|_T}

\def\LMPH#1{\left.#1\mkern1mu\right|_{\mph\Lambda}}

\def\barexymunu{x,\mu\leftrightarrow y,\nu}

\def\us{u_s}
\def\ous{\overline{u}_s}

\def\B#1{\!\left(#1\right)}
\def\BB#1{\!\left[#1\right]}

\def\LARA#1{\!\left\la#1\right\ra}

\def\N#1{:\!#1\!:}

\def\LN{:\!}
\def\RN{\!:}

\def\exval#1#2{\la#1\ra_{#2}}
\def\expval#1#2#3{\la#1\ra_{#2}^{#3}}

\def\trr#1{\text{Tr}\!\left[#1\right]}

\def\be{\begin{equation}}
\def\ee{\end{equation}}

\def\bee{\begin{equation*}}
\def\eee{\end{equation*}}

\def\bg{\begin{equation}\begin{gathered}}
\def\eg{\end{gathered}\end{equation}}

\def\bgg{\begin{equation*}\begin{gathered}}
\def\egg{\end{gathered}\end{equation*}}

\def\ba{\begin{equation}\begin{aligned}}
\def\ea{\end{aligned}\end{equation}}

\def\mph{\lambda}

\def\mphtoL{(\mph\to\Lambda)}

\def\ltoL{(\lambda\to\Lambda)}

\def\izero{\ii0}

\def\sz{s_z}
\def\spinz{\sz\delta^{i3}}

\def\sp{\text{spin\textbullet}}

\def\orb{\text{orb\textbullet}}

\def\spel{\text{spin$\thicksim$}}
\def\orbel{\text{orb$\thicksim$}}

\def\Opr{\boldsymbol{\Omega}s}
\def\Oprim{|\Opr\ra}

\def\Vol{V}

\def\for{\ \text{for} \ }

\def\pref{0.5}

\makeatletter
\newcommand{\pushright}[1]{\ifmeasuring@#1\else\omit\hfill$\displaystyle#1$\fi\ignorespaces}
\newcommand{\pushleft}[1]{\ifmeasuring@#1\else\omit$\displaystyle#1$\hfill\fi\ignorespaces}
\makeatother

\def\Xint#1{\mathchoice
{\XXint\displaystyle\textstyle{#1}}%
{\XXint\textstyle\scriptstyle{#1}}%
{\XXint\scriptstyle\scriptscriptstyle{#1}}%
{\XXint\scriptscriptstyle\scriptscriptstyle{#1}}%
\!\int}
\def\XXint#1#2#3{{\setbox0=\hbox{$#1{#2#3}{\int}$}
\vcenter{\hbox{$#2#3$}}\kern-.5\wd0}}

\def\dashint{\Xint-}

\newcommand{\Rzymskie}[1]{%
  \textup{\uppercase\expandafter{\romannumeral#1}}%
}

\usepackage[ddmmyyyy]{datetime}
\makeatletter
\def\Dated@name{}
\makeatother

\begin{document}

\title{Angular momentum  of the electron: One-loop studies}
\author{Bogdan Damski} 
\affiliation{Jagiellonian University,  Institute of Theoretical Physics, {\L}ojasiewicza 11, 30-348 Krak\'ow, Poland}
\begin{abstract}
We combine bare perturbation theory with the imaginary time evolution technique to 
 study one-loop radiative corrections to  various  components of 
 angular momentum   of the electron.   Our  investigations  are 
based on  the
canonical decomposition of angular momentum, where  spin and orbital components,
associated with  fermionic and electromagnetic degrees of
 freedom, are individually approached.
We use for this purpose quantum
electrodynamics in  the general  covariant  gauge and 
develop a formalism, based on the repeated use of the
Sochocki-Plemelj formula, for  proper enforcement  of the imaginary time
limit. It is then shown that  careful 
implementation of  imaginary time evolutions
is crucial for getting  a correct result for total  angular momentum of the electron 
in the bare perturbative expansion. We also analyze
applicability of the Pauli-Villars regularization 
to our problem,  developing a variant of this technique based on modifications of 
studied observables by subtraction of their  ghost operator counterparts. It is then shown that such an approach 
 leads to the consistent regularization of all angular momenta that we compute.
\end{abstract}
\maketitle

\section{Introduction}
\label{Intro_sec}
The electron,  undoubtedly one of the most fundamental  constituents of matter, is  characterized by a set
of physical properties such as the mass, charge,  magnetic moment, and spin. 

Experimental studies of its mass and charge, $m$ and $e$ below,
 started in the late nineteenth century in a series of experiments conducted  by Thomson
\cite{Thomson1897b}. They have been successfully continued 
ever since. 
By contrast, progress in theoretical characterization  of these parameters is rather
uninspiring,  if we notice that dimensionless  quantities  involving 
them--such as the fine structure constant or ratios of the electron mass to other
lepton  masses--have  never  been convincingly estimated.

The electron's intrinsic magnetic moment was  introduced by Uhlenbeck and Goudsmit \cite{Uhlenbeck1926}
about a century ago in an attempt to explain the 
anomalous Zeeman effect, which was discovered by Preston at the same time
Thomson conducted his electron experiments \cite{Preston1898}. Its understanding rapidly progressed soon after
thanks to   Dirac   \cite{Dirac1928}, whose theory predicted
\be
\frac{e}{2m}
\label{Dirac}
\ee
for  the electron's  magnetic moment. 
Two decades later \cite{SchwingerPR1948},  Schwinger found a more accurate
approximation through  a perturbative   quantum electrodynamics (QED)
calculation replacing (\ref{Dirac}) with
\be
\frac{e}{2m}\B{1+\frac{\alpha}{2\pi}},
\label{Schw}
\ee
where 
\be
\alpha=\frac{e^2}{4\pi}
\label{alfa}
\ee
is the fine structure constant written here  in the Heaviside-Lorentz system of units
combined with $\hbar=c=1$ (we use such units throughout this work).
This prediction immediately explained  spectroscopic ``anomalies'' found  in 
measurements of  Nafe and  Nelson   \cite{NafeNelson1948} 
and  Foley and  Kusch \cite{FoleyKusch1948}
that were done concurrently with Schwinger's  calculations.
Ever since  perturbative calculations of the electron's magnetic moment 
have gone hand in hand with various  experimental measurements 
reaching astonishing accuracy \cite{ComminsAnnRev2012}. These efforts 
allowed for  some of the most stringent tests of QED.

The electron's  spin was introduced 
together with its intrinsic magnetic moment in  \cite{Uhlenbeck1926}. 
It was then put on a firm
theoretical basis by Dirac \cite{Dirac1928}, whose relativistic quantum
mechanics  leads to   the following expression for the  angular momentum  operator  \cite{Greiner}
\be
 \frac{1}{2}\int \d{z}\N{\psi^\dag\Sigma^i\psi}-\ii
 \int \d{z}\varepsilon^{imn}z^m\N{\psi^\dag\partial_n\psi},  
\label{JDirac}  
\ee
where $\psi$ is the Dirac  field operator, $\N{ \ }$ denotes  normal ordering, 
\be
\Sigma^i=\ii\varepsilon^{imn}\gamma^m\gamma^n/2,
\ee
and $\gamma$ are Dirac matrices. 
The first (second) operator in  (\ref{JDirac})
is the fermionic spin (orbital)  angular momentum operator. Consider now  the
electron at rest, whose spin is polarized in the $\pm z$ direction.
The expectation value of operator (\ref{JDirac}), in the corresponding  quantum
state $|\Psi\ra$, is $\spinz$, where 
\be
\sz=\pm\frac{1}{2}
\ee
reflects the fact that the electron's spin equals  one-half. 
The orbital component of the angular momentum operator does not
contribute to such an expectation value 
\be
\LARA{-\ii
 \int \d{z}\varepsilon^{imn}z^m\N{\psi^\dag\partial_n\psi}}_\Psi=0,
\label{Diraco}
\ee
and so one finds 
\be
\LARA{ \frac{1}{2}\int \d{z}\N{\psi^\dag\Sigma^i\psi}}_\Psi=\spinz.
\label{Diracs}
\ee

The situation is considerably more complex in QED, where 
the total angular momentum operator is built of  not only fermionic but also  electromagnetic 
  operators.
The question of how one can attribute angular momentum to different degrees of freedom
is non-trivial  and it  lead to the so-called
angular momentum controversy involving 
 various issues such as the lack of gauge invariant
definition of spin and orbital angular momentum of photons 
and the question of  experimental  relevance of gauge non-invariant quantities \cite{LeaderPhysRep2014}.
More importantly,  in the context of this work, 
 all components of  total angular momentum   of the electron 
receive radiative
corrections \cite{BurkardtPRD2009,LiuPRD2015,JiPRD2016,BDfield}.

The interest in angular momentum decompositions of the electron in particular
and other subatomic particles in general comes from the fact that they
provide  fundamental insights into  properties  of these particles. 
This statement is perhaps best illustrated by experimental and theoretical 
studies of  angular momentum decompositions of nucleons
 performed  over  last four decades and comprehensively summarized in \cite{Deur2019}.

It is the purpose of this work to compute radiative corrections to  right-hand sides of  (\ref{Diraco}) and 
(\ref{Diracs}) as well as to remaining components of total angular momentum
of the electron.
Similar studies were  performed not long ago \cite{LiuPRD2015,JiPRD2016}.
These calculations were done in the light-front formalism,  employed 
the light-cone gauge, and used renormalized perturbation theory. 
They are, on the technical level, very different from  our studies as 
we use imaginary time evolution formalism, work in the general  covariant gauge, and 
employ bare perturbation theory.
Therefore, we  see our work as complementary to  previous efforts.
Among other things, this paper   
discusses   non-trivial results on
implementation of imaginary time evolutions, it presents 
gauge non-invariant angular momenta  from the covariant-gauge perspective, and it conclusively describes
intricacies of proper application  of the Pauli-Villars regularization
to the studied problem. 
Its  outline  is the following. 

We explain in Sec. \ref{Basics_sec} the
approach that we use to carry out  computations.
Next, we describe in Sec. \ref{Fer_spin_sec}  different contributions 
to fermionic spin angular momentum of the electron. 
Remaining angular
momenta--fermionic orbital, electromagnetic spin and orbital, and gauge-fixing ones--are discussed 
in Sec. \ref{Other_sec}.
Then, a proper way of imposing the Pauli-Villars regularization onto  all these expressions
is presented in Sec. \ref{PVV_sec}. One-loop  radiative corrections are computed in Sec. \ref{Regularized_sec}.
The discussion of  obtained results  is presented in Sec. \ref{Discussion_sec}.
Several appendices are added to this paper to make its main body better
readable and to facilitate verification of our calculations. We explain our notation in Appendix \ref{Conventions_sec}
and  collect all  bispinor matrix elements in 
Appendix \ref{Matrix_sec}. 
Intricacies  associated with implementation of imaginary time evolutions
are discussed in Appendix \ref{Implementation_sec}, while 
adaptation  of the Pauli-Villars regularization
technique to our problem is presented in Appendix \ref{Pauli_sec}.
Finally, some integrals from Sec. \ref{Regularized_fer_spin} are evaluated in Appendix
\ref{Integrals_app}.

\section{Basics}
\label{Basics_sec}

The starting point for our considerations is the QED Lagrangian density
\cite{Greiner}
\be
\begin{aligned}
{\cal L} =& -\frac{1}{4} F_{\mu\nu}F^{\mu\nu}+
\frac{\lambda^2}{2}A_\mu A^\mu-\frac{\xi}{2}\B{\partial_\mu A^\mu}^2 + \overline{\psi}\B{\ii\gamma^\mu\partial_\mu-
\mo}\psi- \eo\overline{\psi}\gamma^\mu\psi A_\mu,
\end{aligned}
\label{LL}
\ee
where the second term   is employed to regulate the infrared (IR) sector of the theory,
 while the third one, the so-called gauge-fixing term,  facilitates quantization of the electromagnetic field 
 in the general covariant gauge
 (the term general refers to the arbitrary  greater than zero value of $\xi$). 
The  bare mass and charge of the electron  are  denoted by $\mo$ and $\eo$, 
the photon mass is written as $\lambda$, 
and remaining symbols follow all standard conventions (Appendix
\ref{Conventions_sec}).

We compute total angular momentum through the formula from Sec. 2.4 of \cite{Greiner}
\be
J^i=\frac{1}{2}\varepsilon^{imn}\int \d{z} M^{0mn},
\label{Js}
\ee
where the canonical angular momentum tensor density is given by the
following 
sum of the orbital term, expressed through the canonical energy-momentum tensor density
$\vartheta^{\mu\nu}$, and the spin term
\begin{subequations}
\be
M^{\mu\nu\lambda} = \vartheta^{\mu\lambda}z^\nu -\vartheta^{\mu\nu}z^\lambda +\delta M^{\mu\nu\lambda},
\ee
\be
\begin{aligned}
\label{emtensor}
\vartheta^{\mu\nu}=&\frac{\partial{\cal L}}{\partial(\partial_\mu A^\sigma)}\partial^\nu A^\sigma 
                    +\frac{\partial{\cal L}}{\partial(\partial_\mu \psi)}\partial^\nu \psi
- \eta^{\mu\nu}{\cal L}\\
=&-F^{\mu\sigma} \partial^\nu A_\sigma -\xi\partial_\sigma A^\sigma \partial^\nu A^\mu 
+\ii\overline{\psi}\gamma^\mu\partial^\nu\psi  - \eta^{\mu\nu}{\cal L},
\end{aligned}
\ee

\be
\begin{aligned}
\delta M^{\mu\nu\lambda}=&\frac{\partial{\cal L}}{\partial(\partial_\mu A^\sigma)}
                          \B{\eta^{\nu\sigma}\eta^{\lambda\rho}-\eta^{\lambda\sigma}\eta^{\nu\rho} }  A_\rho
                         +\frac{\partial{\cal L}}{\partial(\partial_\mu\psi)}
                         \frac{1}{4}[\gamma^\nu,\gamma^\lambda]\psi\\
=&F^{\mu\lambda}A^\nu-F^{\mu\nu}A^\lambda+\xi\partial_\sigma A^\sigma
(\eta^{\mu\lambda}A^\nu-\eta^{\mu\nu}A^\lambda) + \frac{\ii}{4}\overline{\psi}\gamma^\mu[\gamma^\nu,\gamma^\lambda]\psi,
\end{aligned}
\ee
\end{subequations}
where $[\,,]$ stands for the commutator. These expressions  lead to 
\be
\begin{aligned}
J^i=& \frac{1}{2}\int \d{z}\psi^\dag\Sigma^i\psi 
-\ii \int \d{z}\varepsilon^{imn}z^m \psi^\dag \partial_n\psi 
   + \int \d{z}  \varepsilon^{imn} F_{m0}A_n\\
   +& \int \d{z}  \varepsilon^{imn} z^m F_{j0}\partial_n A_j 
   + \xi\int \d{z}  \varepsilon^{imn}  z^m\partial_\sigma A^\sigma\partial_n A_0.
\end{aligned}
\label{Jclass}
\ee

First two terms in (\ref{Jclass}), fermionic spin and orbital angular momenta,
have already been introduced in Sec. \ref{Intro_sec}.
The third and fourth term are known as  
electromagnetic
spin and orbital angular momenta. Finally, we  will refer to the last term of (\ref{Jclass}) as 
gauge-fixing angular momentum because it originates from 
the gauge-fixing term in (\ref{LL}). Such a  term is a  unique feature of the covariant gauge approach, and 
so it is quite interesting to see how it contributes to total angular
momentum of the  electron.

The sum of  first four expressions in  (\ref{Jclass})  is  known as the 
Jaffe-Manohar decomposition of  total angular momentum \cite{JaffeNPB1990,LeaderPhysRep2014}. 
As we have shown above,  it follows directly from 
the canonical formalism, which makes it quite 
distinctive. Such a decomposition, however, is not
unique as one can try to modify the density of angular momentum  through either
 Euler-Lagrange equations or through addition of $3$-divergence terms.
Since advantages and disadvantages of different angular momentum decompositions are
comprehensively discussed in \cite{LeaderPhysRep2014}, we will not dwell on
them.

Angular momentum operators are now obtained by replacing  classical
fields  in (\ref{Jclass}) with Heisenberg-picture 
operators and by imposing normal ordering.
In the form suitable for perturbative calculations,  we write  them as
\begin{align}
&J^i_\sp = \int \d{z}\N{\overline{\psi}\,\Gamma^i\psi}, \ \Gamma^i=\frac{\ii}{4}\varepsilon^{imn}\gamma^0\gamma^m\gamma^n,
\label{Jspin}\\
&J^i_\orb= \int \d{z}\N{\overline{\psi}\,\nabla^i_\z\psi},  \ 
\nabla^i_\z=-\ii\gamma^0\varepsilon^{imn}z^m\frac{\partial}{\partial z^n},
\label{Jorb}\\
&J^i_\spel=\int \d{z}  \varepsilon^{imn} \N{F_{m0}A_n},
\label{Jspinel}\\
&J^i_\orbel=\int \d{z}  \varepsilon^{imn} z^m
\N{F_{j0}\partial_n A_j},
\label{Jorbel}\\
&J^i_\xi=\xi \int \d{z}  \varepsilon^{imn} z^m \N{\partial_\sigma A^\sigma\partial_n A_0},
\label{Jxi}
\end{align}
where we have used the bullet $\bullet$    and the wavy line $\thicksim$ to distinguish
fermionic operators from  electromagnetic ones.
The total angular momentum operator is then 
\be
J^i=J^i_\sp+J^i_\orb+J^i_\spel+J^i_\orbel+J^i_\xi.
\label{totalJ}
\ee

We will compute expectation values of operators (\ref{Jspin})--(\ref{Jxi}) 
in the QED ground state with one net electron,\footnote{The term net refers to the fact that besides 
electrons in vacuum electron-positron pairs,
there is  one electron in such a state.} which we denote as $\Oprim$.
As such quantities are time-independent, we set
\be
z=(0,\z)
\label{z0}
\ee
to simplify the discussion  in  intermediate steps (as a self-consistency
check, we have verified that $z^0$ eventually drops out from all expectation values  if it is not set to zero).
Calculations will be performed in the framework of bare perturbation theory 
combined with the imaginary time evolution technique.

Imaginary time evolutions    start from the one-electron ground state of the free Hamiltonian
\be
|\0 s\ra=a^\dag_{\0 s}|0\ra,
\label{0s}
\ee
where $|0\ra$ is the vacuum state of the free theory   and  the operator $a_{\0 s}$
is introduced in Appendix \ref{Conventions_sec}.   
Such a state  describes the  electron at rest  whose  spin is polarized such
that $\exval{J^i}{\0s}=\spinz$ (the same polarization has been employed in Sec.
\ref{Intro_sec}).
Its $4$-momentum
\be
f=(\mo,\0)
\label{f0}
\ee 
frequently appears in the following discussion.
State (\ref{0s})  is then evolved in  time (its non-trivial dynamics is induced by the interaction Hamiltonian
$\int\d{x}{\cal H}_\IN$). Enforcement of the imaginary time limit leads to  \cite{PS}
\begin{subequations}
\begin{align}
\label{fghj_a}
&\exval{\J_\chi}{\Opr}=\limT\expval{\J_\chi}{\Opr}{\LT}, \\ 
&\expval{\J_\chi}{\Opr}{\LT}=
\frac{\la\0s|\T\J^I_\chi\exp(-\ii\intT\dd{x}{\cal H}^I_\IN)|\0s\ra}{
\la\0s|\T\exp(-\ii\intT\dd{x}{\cal H}^I_\IN)|\0s\ra}, \label{fghj_b}\\
&\intT\dd{x}=\intTT dx^0\int\d{x},\\
&\chi=\sp, \orb, \spel, \orbel, \xi,
\label{chi_list}
\end{align}
\label{fghj}%
\end{subequations}
where interaction-picture operators are labeled with the index $I$,
$\J^I_\chi$ operators are obtained by replacing Heisenberg-picture fields 
with their interaction-picture counterparts,\footnote{This may be less obvious
for operators  involving time derivatives of the $4$-potential $A_\mu$--$J^i_\spel$, 
$J^i_\orbel$, and $J^i_\xi$--but it can be  proven there as
well (see e.g. \cite{BDfield}).}
\be
{\cal H}^{I}_\IN= \eo \N{\overline{\psi}_I\gamma^\mu\psi_I}A^{I}_\mu,
\label{Hint}
\ee
and $\T$ is the time-ordering operator.

To proceed with  (\ref{fghj}), we will  need fermionic  
\be
S(x-y)=
\contraction{}{\psi}{_{I}(x)}{\psi}\psi_{I}(x)\overline{\psi}_{I}(y)=
\la0|\T \psi_I(x)\overline{\psi}_I(y)|0\ra=
\ii\int\frac{\dd{p}}{(2\pi)^4}\frac{\gamma\cdot p+\mo}{p^2-\mo^2+\izero}e^{-\ii p\cdot (x-y)}
\label{prop_fer}
\ee
and electromagnetic 
\begin{subequations}
\begin{align}
&D_{\mu\nu}(x-y)=
\contraction{}{A}{^{I}_\mu(x)}{A}A^{I}_\mu(x)A^{I}_\nu(y)=
\la0|\T A^I_\mu(x)A^I_\nu(y)|0\ra=
-\ii\int\frac{\dd{p}}{(2\pi)^4}\frac{d_{\mu\nu}(p)}{p^2-\lambda^2+\izero}e^{-\ii p\cdot (x-y)},\\
&d_{\mu\nu}(p)= \eta_{\mu\nu}+\xxii\frac{p_\mu p_\nu}{p^2-\lambda^2/\xi+\izero}
\label{prop_mod}%
\end{align}
\label{prop_el}%
\end{subequations}
propagators.   The former expression is given by the standard formula, while the latter one can be 
either derived using the 
trick from Sec. 7.6 of \cite{Greiner} or taken from Sec. 33.4 of \cite{ColemanBook}.

Evaluation of (\ref{fghj_b})  will be performed  with $T>0$ and then  the limit 
\be
\barelimT
\label{limitT}
\ee
will be taken. Proper evaluation of this limit is  no trivial matter
in some of our computations. To illustrate the subtle point here, we note that 
integration over time in (\ref{fghj_b}) leads to expressions of the form 
\be
\intTT \frac{dx^0}{2\pi}  e^{\ii x^0 P^0}=\frac{\sin(TP^0)}{\pi P^0},
\label{dx0sin}
\ee
where $P^0$ is some combination of timelike components of $4$-momenta. 
Limit (\ref{limitT}) cannot be taken on (\ref{dx0sin}). 
The standard textbook solution of this complication
 is to transfer the $-\izero$ from the limit   to the
imaginary part of  propagators' denominators (see e.g. Sec. 4.4 of \cite{PS}). After that, the limit 
$T\to\infty$ is taken. This  leads to the Dirac delta function
due to the following well-known identity
\be
\delta(P^0)=\lim_{T\to\infty}\frac{\sin(TP^0)}{\pi P^0}.
\label{deltac}
\ee
Such a procedure presumably greatly simplifies calculations. However, it
leads to the incorrect result for fermionic spin angular momentum of the electron
and it actually complicates a bit the discussion of its fermionic orbital angular
momentum.
Therefore,  a more rigorous approach is needed and we develop it in Appendix
\ref{Implementation_sec}. Among other things, such an approach can be used for
showing that the above-mentioned heuristic procedure provides correct results
for other angular momenta that we discuss.

Next, we note that due to the commutation of the total angular momentum operator
with the Hamiltonian, angular momentum in  states $|\0 s\ra$ and $\Oprim$ is the same, consequently
\be
\exval{J^i}{\Opr}=\spinz.
\label{toot}
\ee
Furthermore, the expectation value of each individual angular momentum operator, say $J^i_\chi$
with $\chi$ given by (\ref{chi_list}), must be either  directly proportional to $\spinz$ or vanish.
It is so because after averaging over spin projections of the electron,
there is no  preferred direction in the three-dimensional space, 
where  $\exval{J^i_\chi}{\Opr}$ is discussed. Hence,  $\exval{J^i_\chi}{\Opr}$
cannot have  the $\sz$-independent component.\footnote{As a self-consistency check, 
we have directly verified for $\xi=1$ that this is indeed the case in  all our calculations. 
The  same explicit verification has been also performed for expectation values of ghost operators
$\tilde J^i_\chi$ discussed in Appendix \ref{Pauli_sec}.}
We will use this observation over and over again to simplify  
calculations.

Moreover, since we will be doing  the perturbative expansion around the one-electron state,
we will be encountering the normalizing constant
\be
\Vol=\la\0s|\0s\ra=\int \frac{\d{x}}{(2\pi)^3}.
\label{Vfact}
\ee
While such a constant  is formally infinite,  it gets unambiguously  cancelled  during 
computations. This happens because all expressions that contribute to the final result describe processes
that happen homogeneously in space. As a result, the outermost spatial integral in every such expression 
is done over a function that is  constant in space, and so it   exactly cancels down normalizing constant 
(\ref{Vfact}) appearing in the denominator of such an expression. Needless to
say, factors like (\ref{Vfact}) are frequently encountered in studies
involving delocalized states
(see e.g. above-cited \cite{LeaderPhysRep2014}).

We also mention that we will draw position-space Feynman diagrams in Figs.
\ref{0th_order}--\ref{2nd_order_orbital} to illustrate contributions to
fermionic spin and orbital angular momenta of the electron. 
The diagram from Fig. X    will be referred to as Diag. X. 
Rules for drawing these diagrams can be deduced without much effort   by 
comparing them to  analytical expressions that we list for them.
There is no need to linger  over these rules because 
all diagrams will be drawn only  after  analytical expressions will be
worked out.

Finally, for the sake of brevity,  we will drop the term
\be
O(\eo^4)
\ee
from all expressions for expectation values of  angular
momentum operators.

\section{Perturbative expansion for fermionic spin angular momentum}
\label{Fer_spin_sec}
We will derive here the IR-regularized expression for fermionic spin angular momentum of
the electron. 
To proceed, we expand (\ref{fghj_b}) in the series in $\eo$ 

\begin{subequations}
\begin{align}
\expval{\J_\sp}{\Opr}{\LT}
&= \frac{\la\0s|\J^I_\sp|\0s\ra}{\Vol} \label{Jpert1} \\
&-\frac{1}{2}
\frac{\la\0s|\T\J^I_\sp 
\intT\dd{x}{\cal H}^{I}_\IN\intT\dd{y}{\cal H}^{I}_\IN|\0s\ra}{\Vol} \label{Jpert2}\\
&+\frac{1}{2}
\frac{\la\0s|\J^I_\sp|\0s\ra}{\Vol}
\frac{\la\0s|\T\intT\dd{x}{\cal H}^{I}_\IN\intT\dd{y}{\cal H}^{I}_\IN|\0s\ra}{\Vol}. \label{Jpert3} 
\end{align}
\label{pertLT}%
\end{subequations}

\begin{figure}[t]
\includegraphics[width=0.35\columnwidth]{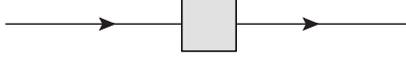}
\caption{Diagrammatic illustration of $\ous\Gamma^i\us/(2\pi)^3$ from 
(\ref{Jpert1_prim}).
The grey box  stands for  the operator $\Gamma^i$ from (\ref{Jspin}). External lines  are for
zero-momentum electrons (the same notation is used in all our figures).
}
\label{0th_order}
\end{figure}

Zeroth-order contribution (\ref{Jpert1}) is  illustrated in Fig. \ref{0th_order}. We obtain after using 
(\ref{ext_contractions}) and (\ref{u})
\be
\frac{\la\0s|(\J^I_\sp)^i|\0s\ra}{\Vol}=\frac{\ous\Gamma^i\us}{\Vol}\int\ddd{z}
=\spinz.
\label{Jpert1_prim}
\ee

\begin{figure}[t]
\includegraphics[width=\pref\columnwidth, clip=true]{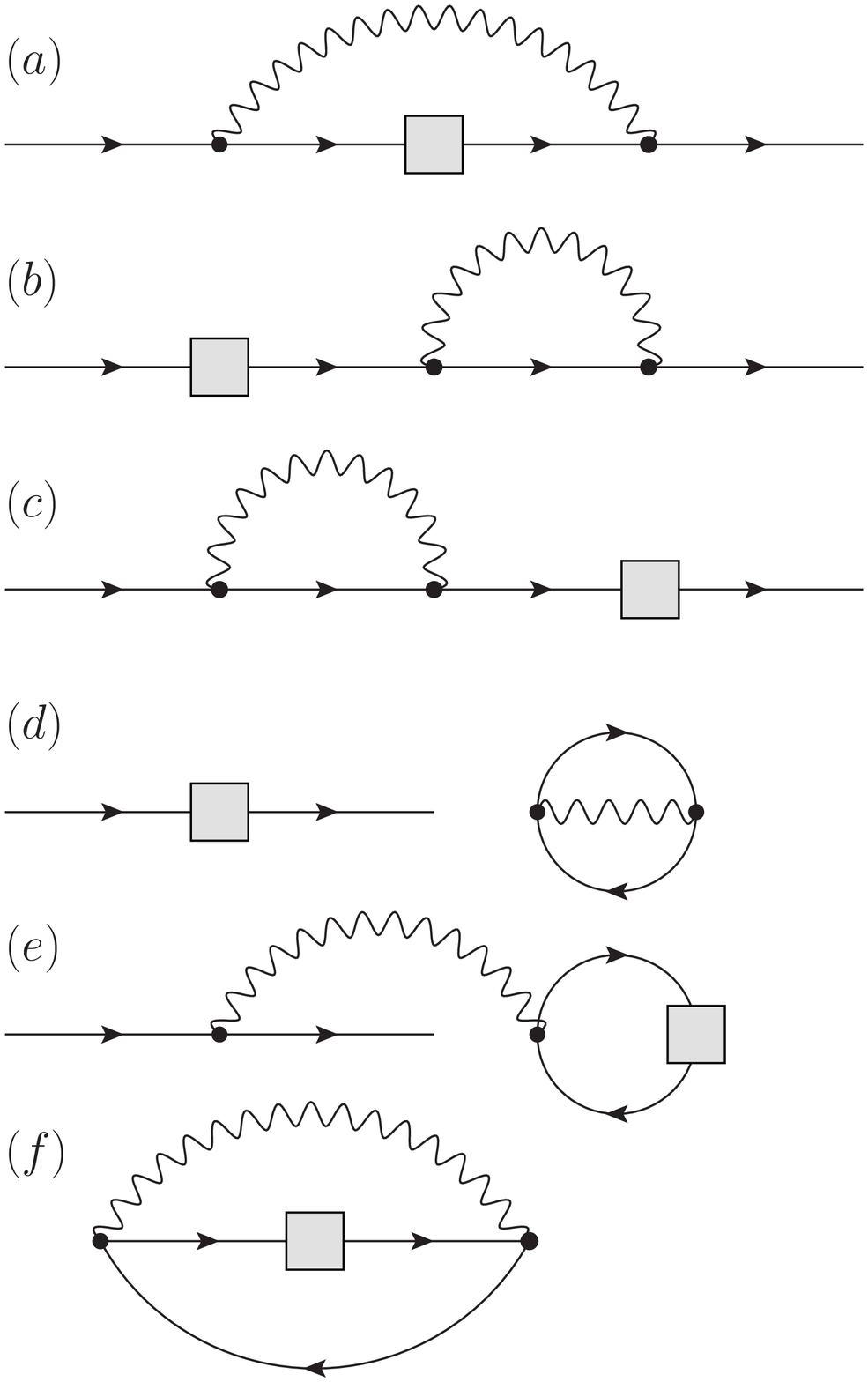}
\caption{The (a)--(f) panels illustrate  photon-propagator contractions with
expressions (\ref{q1})--(\ref{q6}), respectively. }
\label{2nd_order_licznik}
\end{figure}

To compute (\ref{Jpert2}), we need the following matrix  element 
that can be obtained through Wick's theorem combined  with  (\ref{ext_contractions})
\begin{subequations}
\begin{align}
\nonumber
\la\0s|\T\N{\overline{\psi}_I(z)&\Gamma^i\psi_I(z)}
\N{\overline{\psi}_I(x)\gamma^\mu\psi_I(x)}
\N{\overline{\psi}_I(y)\gamma^\nu\psi_I(y)}      |\0s\ra =\\
& \frac{e^{\ii f\cdot (x-y)}}{(2\pi)^3} \ous\gamma^\mu S(x-z)\Gamma^i S(z-y)\gamma^\nu\us
\label{q1} \\
+&\frac{e^{\ii f\cdot (z-y)}}{(2\pi)^3}\ous\Gamma^i S(z-x)\gamma^\mu S(x-y)\gamma^\nu \us
\label{q2} \\
+&\frac{e^{\ii f\cdot (x-z)}}{(2\pi)^3}\ous\gamma^\mu S(x-y)\gamma^\nu S(y-z)\Gamma^i \us
\label{q3} \\
-&\frac{1}{2(2\pi)^3} 
\trr{S(y-x)\gamma^\mu S(x-y)\gamma^\nu}\ous\Gamma^i \us
\label{q4}\\
-&\frac{1}{(2\pi)^3}
\trr{S(y-z)\Gamma^i S(z-y)\gamma^\nu}\ous\gamma^\mu \us
\label{q5}\\
-&\Vol \trr{S(x-z)\Gamma^iS(z-y)\gamma^\nu S(y-x)\gamma^\mu} 
\label{q6}\\
+&(\barexymunu \ \text{on all terms}).
\label{q7}
\end{align}
\label{q}%
\end{subequations}
Matrix element (\ref{q}) can be additionally simplified with (\ref{z0}) and (\ref{f0}) leading to
$e^{\ii f\cdot z}=1$.
Its contractions  with the photon propagator are diagrammatically depicted in Fig. \ref{2nd_order_licznik}.

\begin{figure}[t]
\includegraphics[width=0.55\columnwidth]{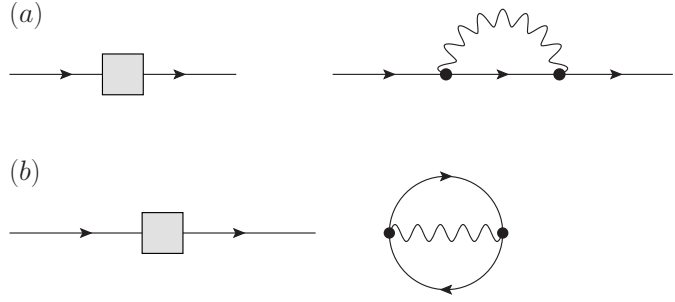}
\caption{The (a) and (b) panels illustrate  photon-propagator contractions with
expressions (\ref{qq1}) and (\ref{qq2}), respectively. }
\label{2nd_order_mianownik}
\end{figure}

To compute (\ref{Jpert3}), we proceed similarly as in (\ref{q}) getting
\begin{subequations}
\begin{align}
&\frac{\la\0s|\N{\overline{\psi}_I(z)\Gamma^i\psi_I(z)}|\0s\ra}{\Vol} 
\la\0s|\T\N{\overline{\psi}_I(x)\gamma^\mu\psi_I(x)}
\N{\overline{\psi}_I(y)\gamma^\nu\psi_I(y)}
|\0s\ra=\nonumber \\
&\phantom{+}\frac{e^{\ii f\cdot (x-y)}}{(2\pi)^6}\frac{\ous\Gamma^i \us}{\Vol}
\ous\,\gamma^\mu S(x-y)\gamma^\nu \us 
\label{qq1}\\
&-\frac{1}{2(2\pi)^3} 
\trr{S(y-x)\gamma^\mu S(x-y)\gamma^\nu}\ous\Gamma^i \us \label{qq2}\\
&+(\barexymunu \ \text{on all terms}),
\label{qq3}
\end{align}
\label{qq}%
\end{subequations}
whose contractions  with the photon propagator are diagrammatically
shown in Fig. \ref{2nd_order_mianownik}.  
Replacements  (\ref{q7}) and (\ref{qq3}) produce a  factor
of $2$ during evaluation of  diagrams, which   cancels down  a prefactor of
$1/2$
from  (\ref{Jpert2}) and (\ref{Jpert3}).

To correctly evaluate   contributions of different diagrams to  fermionic spin
angular momentum of the electron, one must properly enforce limit (\ref{limitT}).
This has to be carefully  done because the standard procedure outlined between 
(\ref{dx0sin}) and (\ref{deltac})  leads to incorrect
results when Diags. \ref{2nd_order_licznik}b, \ref{2nd_order_licznik}c, and
\ref{2nd_order_mianownik}a are considered. 
The comprehensive discussion of  the appropriate way of handling the imaginary time
limit can be found  in Appendix \ref{Implementation_sec}. We will frequently refer the reader to it
quoting below only its final outcomes for individual diagrams.

Finally,  to make equations a bit more compact, we  introduce the following notation
\begin{align}
\label{tildeq_a}
&\text{Diag. X}=\limT \TT{\text{Diag. X}},\\
\label{tildeq}
&\tilde q=(q^0,\p), \ \bar{k}=(k^0,-\p).
\end{align}
We are now ready to discuss  diagrams.

{\bf Diagram \ref{2nd_order_mianownik}a}.
We start with
\ba
\TT{\text{Diag.  \ref{2nd_order_mianownik}a}}=
\frac{\eo^2}{\Vol^2}\int\ddd{z}\ous\Gamma^i\us  \intT\dd{x}\dd{y} 
\frac{e^{\ii f\cdot (x-y)}}{(2\pi)^3}
D_{\mu\nu}(x-y) 
 \ous\gamma^\mu S(x-y)\gamma^\nu \us\\
=\frac{\eo^2\spinz}{(2\pi)^3\Vol}\intT\dd{x}\dd{y}\int\dddd{p}\dddd{k}
\frac{e^{\ii(f-k-p)\cdot (x-y)}}{k^2-\lambda^2+\izero}
\frac{\ous\gamma^\mu(\gamma\cdot p+\mo)\gamma^\nu \us}{p^2-\mo^2+\izero}
d_{\mu\nu}(k)\\
=2\eo^2\spinz\int\dddd{p}dk^0 F(k^0,p^0)
\frac{\sin^2[T(k^0+p^0-\mo)]}{(k^0+p^0-\mo)^2},
\label{3A}
\ea
where identities (\ref{ubaru2}) and (\ref{xiubaru4}) have been employed to get 
\ba
F(k^0,p^0)=&\frac{2}{\pi}\frac{2\mo-p^0}{(\bar{k}^2-\lambda^2+\izero)(p^2-\mo^2+\izero)} \\ 
+&
\frac{1-\xi}{\pi\xi}\frac{2k^0\bar{k}\cdot p +\bar{k}^2(\mo-p^0)}{
(\bar{k}^2-\lambda^2+\izero)
(\bar{k}^2-\lambda^2/\xi+\izero)
(p^2-\mo^2+\izero)}.
\label{Fcov}
\ea
Note that
we only list  those arguments of the function $F$ that are most relevant for
enforcement of the imaginary time limit.
Using (\ref{chi}), we get 
\begin{subequations}
\begin{align}
\text{Diag.  \ref{2nd_order_mianownik}a}=&
\label{diagram3a_1}
2\pi \eo^2\spinz\int \dddd{p} F(\mo-p^0,p^0) \limT  T \\
\label{diagram3a_2}
+&\frac{\eo^2\spinz}{2} \int \dddd{p}dk^0 
\BB{\frac{F(k^0,p^0)}{(k^0+p^0-\mo+\izero)^2}+ \frac{F(k^0,p^0)}{(k^0+p^0-\mo-\izero)^2}}.
\end{align}
\label{diagram3a}%
\end{subequations}

It is now worth to stress that  the  procedure  described
between (\ref{dx0sin}) and  (\ref{deltac}) produces the following ill-defined factor 
under the integral sign
\be
[\delta(k^0+p^0-\mo)]^2,
\ee
which gives a warning sign that such a simplification  is meaningless in this case. By ignoring 
this fact, one ends with term  (\ref{diagram3a_1})
after a formal identification of $\delta(0)$ with 
\be
\limT\int_{-T}^T \frac{dx^0}{2\pi}.
\ee
Leaving aside the discussion of this dubious substitution, 
such a procedure  {\it misses} crucially-important term 
 (\ref{diagram3a_2}), whose derivation requires a more sophisticated 
analytical approach (Appendix \ref{Implementation_sec}). 
We also mention that 
  (\ref{diagram3a_1}) cancels out with similar terms from Diags. 
\ref{2nd_order_licznik}b and \ref{2nd_order_licznik}c.\footnote{The sum of
Diags.  \ref{2nd_order_licznik}b, \ref{2nd_order_licznik}c, and
\ref{2nd_order_mianownik}a is entirely determined by careful enforcement 
of limit (\ref{limitT}). It cannot be obtained by the simplified procedure 
mentioned between (\ref{dx0sin}) and  (\ref{deltac}).}

{\bf Diagrams \ref{2nd_order_licznik}b and \ref{2nd_order_licznik}c}. 
Now, we compute 
\ba
\TT{\text{Diag.  \ref{2nd_order_licznik}b}}=-\frac{\eo^2}{\Vol} \intT\dd{x}\dd{y}\int\ddd{z}
e^{\ii f\cdot (z-y)}
D_{\mu\nu}(x-y)
\ous\Gamma^i S(z-x)\gamma^\mu S(x-y)\gamma^\nu \us\\
=-\frac{\ii\eo^2}{\Vol} \intT\dd{x}\dd{y}\int\ddd{z}\int \dddd{k} \dddd{p} \dddd{q}
\frac{e^{\ii (q-k-p)\cdot x
+\ii (k+p-f)\cdot y
+\ii (f-q)\cdot z}}{k^2-\lambda^2+\izero} \\
\cdot
\frac{\ous\Gamma^i(\gamma\cdot q+\mo)\gamma^\mu (\gamma\cdot p+\mo)\gamma^\nu \us}{
(q^2-\mo^2+\izero)(p^2-\mo^2+\izero)}d_{\mu\nu}(k).
\label{diagram2b_1}
\ea
 
Employing  (\ref{ubaru4}) and (\ref{xiubaru2}), (\ref{diagram2b_1}) can be written as 
\begin{multline}
\TT{\text{Diag.  \ref{2nd_order_licznik}b}}=
-2\ii\eo^2\spinz \int \dddd{p} \frac{dq^0dk^0}{2\pi}
\frac{F(k^0,p^0)}{q^0-\mo+\izero}\\ 
\cdot\frac{\sin[T(k^0+p^0-q^0)]}{k^0+p^0-q^0} \frac{\sin[T(k^0+p^0-\mo)]}{k^0+p^0-\mo}.
\label{diagram2b_2}
\end{multline}

We now  note that the procedure outlined between (\ref{dx0sin})
and  (\ref{deltac})
 leads to  $\delta(q^0-\mo)$ producing a meaningless factor of $1/\izero$ 
 in the  expression for
 $\text{Diag.  \ref{2nd_order_licznik}b}$. 
This leaves no doubts that careful
 implementation of the imaginary time limit is necessary.

So, using (\ref{2chii}), we find 
\ba
\text{Diag.  \ref{2nd_order_licznik}b}=&-\pi\eo^2\spinz\int \dddd{p}   F(\mo-p^0,p^0) \limT T\\
&-\frac{\eo^2\spinz}{2}  \int \dddd{p} dk^0
\frac{F(k^0,p^0)}{(k^0+p^0-\mo-\izero)^2}.
\label{diagram2b_3}%
\ea

Computation of
\be
\TT{\text{Diag. \ref{2nd_order_licznik}c}}=-\frac{\eo^2}{\Vol} \intT\dd{x}\dd{y}\int\ddd{z}
e^{\ii f\cdot (x-z)}
D_{\mu\nu}(x-y)
\ous\gamma^\mu S(x-y)\gamma^\nu S(y-z)\Gamma^i \us
\label{diagram2c_1}
\ee
follows now straightforwardly as through formal manipulations one can show that   
\ba
\text{Diag. \ref{2nd_order_licznik}c} =\text{Diag. \ref{2nd_order_licznik}b}
\label{diagram2c_2}
\ea
if (\ref{z0}) holds.

{\bf Diagram \ref{2nd_order_licznik}a}.
We compute here
\ba
\TT{\text{Diag.  \ref{2nd_order_licznik}a}}=-\frac{\eo^2}{\Vol} \intT\dd{x}\dd{y}\int\ddd{z}
e^{\ii f\cdot (x-y)}
D_{\mu\nu}(x-y)
\ous\gamma^\mu S(x-z)\Gamma^i S(z-y)\gamma^\nu \us\\
=-\frac{\ii\eo^2}{\Vol} \intT\dd{x}\dd{y}\int\ddd{z}\int \dddd{k} \dddd{p} \dddd{q}
\frac{e^{
\ii 
(f-k-p)\cdot x
+\ii (k+q-f)\cdot y
+\ii 
(p-q)\cdot z
}}{k^2-\lambda^2+\izero} \\
\cdot
\frac{\ous\gamma^\mu(\gamma\cdot p+\mo)\Gamma^i (\gamma\cdot q+\mo)\gamma^\nu \us}{(p^2-\mo^2+\izero)(q^2-\mo^2+\izero)}
d_{\mu\nu}(k)\\
=-4\ii\eo^2 \int  \dddd{p}\frac{dk^0 dq^0}{(2\pi)^2}
\frac{\ous\gamma^\mu(\gamma\cdot p+\mo)\Gamma^i (\gamma\cdot\tilde q+\mo)\gamma^\nu \us}{
(\bar{k}^2-\lambda^2+\izero)(p^2-\mo^2+\izero)(\tilde
q^2-\mo^2+\izero)}d_{\mu\nu}(\bar{k})\\
\cdot\frac{\sin[T(k^0+p^0-\mo)]}{k^0+p^0-\mo}\frac{\sin[T(k^0+q^0-\mo)]}{k^0+q^0-\mo}.
\label{diagram2a_1}
\ea
With the help of (\ref{ubaru6}), (\ref{xiubaru6}),  and
(\ref{chiii}) we arrive at 
\begin{multline}
\text{Diag.  \ref{2nd_order_licznik}a}=-\ii\eo^2\spinz\int\dddd{p}
\left[
\frac{2(p^2+\mo^2)+4(p_3)^2}{(p^2-\mo^2+\izero)^2[(p-f)^2-\lambda^2+\izero]}\right.\\
\left.
+\xxii\frac{1}{[(p-f)^2-\lambda^2+\izero][(p-f)^2-\lambda^2/\xi+\izero]}\right].
\label{diagram2a_2}
\end{multline}
We mention in passing that the procedure discussed between (\ref{dx0sin}) and  (\ref{deltac}) gives
a correct result here (no singularities are encountered during its
implementation).

There are no other one-loop  contributions to fermionic spin angular momentum of the electron in covariantly quantized QED. 
Indeed, disconnected vacuum bubble 
Diags. \ref{2nd_order_licznik}d and \ref{2nd_order_mianownik}b immediately cancel out due to the 
difference in overall signs of (\ref{Jpert2}) and (\ref{Jpert3}). 
Therefore, there is no need to write down  expressions for them. 
Moreover,  
\be
\text{Diag.  \ref{2nd_order_licznik}e}=\limT\frac{\eo^2}{\Vol} \intT\dd{x}\dd{y}\int\ddd{z}
D_{\mu\nu}(x-y) \trr{S(y-z)\Gamma^iS(z-y)\gamma^\nu}
\ous\gamma^\mu \us
\label{diagram2e}
\ee
and
\be
\text{Diag.  \ref{2nd_order_licznik}f}=\limT\eo^2 \intT\dd{x}\dd{y}\int\d{z}
D_{\mu\nu}(x-y) \trr{S(x-z)\Gamma^iS(z-y)\gamma^\nu S(y-x)\gamma^\mu}
\label{diagram2f}
\ee
also do not contribute  because they are both $\sz$-independent--see identity 
(\ref{ubaru0})   
and the discussion below (\ref{toot}).

The final IR-regularized result for fermionic spin angular momentum of the electron comes from   
Diags. \ref{0th_order}, \ref{2nd_order_licznik}a--\ref{2nd_order_licznik}c, and \ref{2nd_order_mianownik}a
\be
\expval{J^i_\sp}{\Opr}{\lambda}=
\text{Diag. \ref{0th_order}}+
\text{Diag. \ref{2nd_order_licznik}a}+
\text{Diag. \ref{2nd_order_licznik}b}+
\text{Diag. \ref{2nd_order_licznik}c}+\text{Diag. \ref{2nd_order_mianownik}a},
\label{unJspin}
\ee
where the superscript $\lambda$ indicates the fact that  the IR regularization
is present in (\ref{unJspin}). This expression  
 can be obtained by adding (\ref{Jpert1_prim}) and  (\ref{diagram2a_2}) to 
\begin{multline}
\text{Diag. \ref{2nd_order_licznik}b}+
\text{Diag. \ref{2nd_order_licznik}c}+
\text{Diag. \ref{2nd_order_mianownik}a}=\\
2\ii\eo^2\spinz\int\frac{\dd{p}}{(2\pi)^4}
\BB{
\frac{\om{p}^2(p^2-\mo^2)+\lambda^2[3(p^0-\mo)^2-\om{p}^2]}{\lambda^2(p^2-\mo^2+\izero)[(p-f)^2-\lambda^2+\izero]^2}
-\frac{\om{p}^2+\lambda^2/\xi}{\lambda^2[(p-f)^2-\lambda^2/\xi+\izero]^2}
}.
\label{suma3}
\end{multline}
Note that there is no singularity in the  integrand of (\ref{suma3}) at
$\lambda=0$  despite a factor  of $\lambda^2$  in 
denominators, which   can be shown by rearranging terms.

\section{Perturbative expansion for other angular momenta}
\label{Other_sec}
We will derive here IR-regularized expressions for fermionic orbital angular
momentum, electromagnetic spin and orbital angular momenta, and 
gauge-fixing angular momentum.

Such an expression  for  fermionic orbital angular momentum can be obtained through 
straightforward   modifications of calculations  reported in Sec. \ref{Fer_spin_sec}. 
We will discuss its derivation  in Sec. \ref{Fer_orb_sec}.

Results  for  electromagnetic spin, electromagnetic orbital, and gauge-fixing
angular momenta have to be derived from scratch, which is simplified by the following observation. 
Namely, it can be  easily shown with  (\ref{chiii}), 
that IR-regularized expressions for these angular momenta can be obtained from (\ref{fghj})  through the replacement
\be
\limT\int_T \dd{x}\to\int\dd{x}.
\label{qwwww}
\ee
The hint that such a simplification is going to work  comes from the fact that
(\ref{qwwww}), which amounts to the procedure described between  (\ref{dx0sin}) and  (\ref{deltac}),
does not lead to singular expressions here.
By combining (\ref{qwwww}) with  the following observation 
\be
\la\0s|\J^I_\chi|\0s\ra=\0, \ \ \chi=\spel, \orbel, \xi,
\ee
we find from (\ref{fghj}) that 
\be
\label{Jpert_chia}
\exval{\J_\chi}{\Opr}=-\frac{1}{2\Vol}\int \dd{x}\dd{y}
\la\0s|\T\J^I_\chi {\cal H}^{I}_\IN(x) {\cal H}^{I}_\IN(y)|\0s\ra,
\ee
which will be used in Secs. \ref{El_spin_sec}--\ref{G_fix_sec}.

\subsection{Fermionic orbital  angular momentum}
\label{Fer_orb_sec}
\begin{figure}[t]
\includegraphics[width=0.55\columnwidth, clip=true]{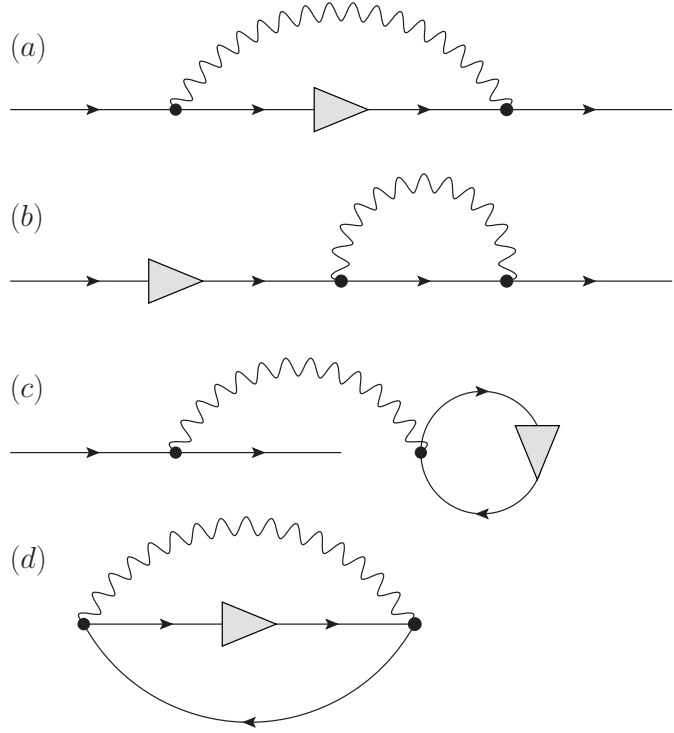}
\caption{The (a)--(d) panels  illustrate   photon-propagator contractions with
expressions (\ref{o1})--(\ref{o4}), respectively. 
The grey triangle stands for the operator $\nabla^i_\z$, which is  defined in (\ref{Jorb}). It
acts on the fermionic propagator attached to its vertex.
}
\label{2nd_order_orbital}
\end{figure}

We begin by noting that
\be
\la\0s|\J^I_\orb|\0s\ra=\0,
\ee
which simplifies a bit the  following discussion based on (\ref{fghj}). 
Another matrix element that we need to know is 
\begin{subequations}
\begin{align}
\nonumber
\la\0s|\T\N{\overline{\psi}_I(z)&\nabla^i_\z\psi_I(z)}
\N{\overline{\psi}_I(x)\gamma^\mu\psi_I(x)}
\N{\overline{\psi}_I(y)\gamma^\nu\psi_I(y)}      |\0s\ra =\\
&\phantom{+}\frac{e^{\ii f\cdot (x-y)}}{(2\pi)^3}\ous\gamma^\mu S(x-z)\nabla_\z^i S(z-y)\gamma^\nu \us
 \label{o1} \\
&+\frac{e^{\ii f\cdot (z-y)}}{(2\pi)^3}\ous\nabla_\z^i S(z-x)\gamma^\mu S(x-y)\gamma^\nu \us
 \label{o2} \\
&-\frac{1}{(2\pi)^3}
\trr{S(y-z)\nabla_\z^i S(z-y)\gamma^\nu} \ous\gamma^\mu \us
 \label{o3}\\
&-\Vol \trr{S(x-z)\nabla_\z^iS(z-y)\gamma^\nu S(y-x)\gamma^\mu} 
\label{o4}\\
&+(\barexymunu \ \text{on all terms}),
\label{o5}
\end{align}
\label{o}%
\end{subequations}
whose contractions with the photon propagator are diagrammatically depicted in Fig. \ref{2nd_order_orbital}. 
Such an expression can be  obtained by replacing $\Gamma^i$ in 
(\ref{q}) by $\nabla^i_\z$ and by noting that the latter operator
gives zero when acting on  bispinors $\us$  (\ref{u}). 
Replacements (\ref{o5}) produce a  factor
of $2$ during evaluation of  diagrams, which cancels down a  prefactor of  $1/2$
coming from the  second order  expansion of the exponential function  in the numerator of
(\ref{fghj_b}).

Armed with (\ref{o}), we can proceed similarly as in Sec. \ref{Fer_spin_sec}
discussing each diagram separately.
We start from the only diagram, which 
yields a non-zero contribution to fermionic orbital angular momentum of the
electron.

{\bf Diagram \ref{2nd_order_orbital}a}. We employ notation (\ref{tildeq_a}) and  compute 
\ba
\TT{\text{Diag. \ref{2nd_order_orbital}a}}=-\frac{\eo^2}{\Vol}\intT\dd{x}\dd{y}\int\ddd{z}
e^{\ii f\cdot (x-y)}D_{\mu\nu}(x-y)\ous\gamma^\mu S(x-z)\nabla_\z^i S(z-y)\gamma^\nu \us\\
=-\ii\frac{\eo^2}{\Vol}\intT\dd{x}\dd{y}\int\ddd{z}\int\dddd{k}\dddd{p}\dddd{q}
\frac{e^{\ii (f-k-p)\cdot x+\ii(k+q-f)\cdot y+\ii (p-q)\cdot z}}{k^2-\lambda^2+\izero}\\
\cdot (\z\times\q)^i
\frac{\ous\gamma^\mu(\gamma\cdot p+\mo)\gamma^0(\gamma\cdot q+\mo)\gamma^\nu\us}{(p^2-\mo^2+\izero)(q^2-\mo^2+\izero)}d_{\mu\nu}(k).
\ea
Next, we use  
\be
\int\ddd{z} 
e^{\ii (p-q)\cdot z}
(\z\times\q)^i
=\varepsilon^{imn}q^n\frac{\ii}{2}
\B{  \frac{\partial}{\partial p^m}  -  \frac{\partial}{\partial q^m}  }\delta(\p-\q),
\label{pq_derz}
\ee
and integrate by parts to move derivatives acting on $\delta(\p-\q)$ to the rest of
the integrand. Boundary terms from integration by parts disappear. For
example,  because the integrand of the resulting  surface integral is
proportional to 
\be
\varepsilon^{imn}q^mq^n=0.
\label{ijn_delta}
\ee
Derivatives of propagators' denominators lead to the same factors and so they also do not contribute.
A similar thing  can be said about derivatives of the exponential  term because 
\be
\int \d{x}\d{y} \delta(\p-\q)
\B{  \frac{\partial}{\partial p^m}  -  \frac{\partial}{\partial q^m}}e^{\ii(q\cdot y-p\cdot x)}
\sim \int \d{x}\d{y} (\x+\y)^m e^{\ii\q\cdot(\x-\y)}=0.
\label{exp_vanish}
\ee
In the end, after spacetime integrations  and employment of  (\ref{chiii}), we arrive at
\be
\text{Diag. \ref{2nd_order_orbital}a}
=\frac{\eo^2}{2}\int\dddd{p}
\frac{
\varepsilon^{imn}p^n\,
\ous\gamma^\mu\{\gamma^m\gamma^0,\gamma\cdot p+\mo\}\gamma^\nu\us}{
(p^2-\mo^2+\izero)^2[(p-f)^2-\lambda^2+\izero]}d_{\mu\nu}(f-p),
\label{4aa}
\ee
where $\{\,,\}$ stands for the anticommutator.

Finally, we use  (\ref{ubaru12}) and (\ref{ubaru12next}) to get 
\begin{multline}
\text{Diag. \ref{2nd_order_orbital}a}
=-4\ii\eo^2\spinz \int\dddd{p} 
\frac{(p_1)^2+(p_2)^2}{(p^2-\mo^2+\izero)^2[(p-f)^2-\lambda^2+\izero]}\\
\cdot\B{1+\xxiidwa\frac{p^2-\mo^2}{(p-f)^2-\lambda^2/\xi+\izero}}.
\label{4a4a}
\end{multline}
It is perhaps worth to mention that the procedure discussed between (\ref{dx0sin}) and  (\ref{deltac}) 
leads to the same result for this diagram.

{\bf Diagram \ref{2nd_order_orbital}b}. We study now
\ba
\TT{\text{Diag. \ref{2nd_order_orbital}b}}=-\frac{\eo^2}{\Vol}\intT\dd{x}\dd{y}\int\ddd{z}
e^{\ii f\cdot (z-y)}D_{\mu\nu}(x-y)\ous\nabla_\z^i S(z-x)\gamma^\mu S(x-y)\gamma^\nu \us\\
=-\ii\frac{\eo^2}{\Vol}\intT\dd{x}\dd{y}\int\ddd{z}\int\dddd{k}\dddd{p}\dddd{q}
\frac{e^{\ii(q-k-p)\cdot x+\ii(k+p-f)\cdot y+\ii(f-q)\cdot z}}{k^2-\lambda^2+\izero}\\
\cdot (\z\times\q)^i
\frac{\ous\gamma^0(\gamma\cdot q+\mo)\gamma^\mu(\gamma\cdot p+\mo)\gamma^\nu \us}
{(q^2-\mo^2+\izero)(p^2-\mo^2+\izero)}d_{\mu\nu}(k).
\label{xdse}
\ea
Next, we note that 
\be
\int\ddd{z} 
e^{\ii (f-q)\cdot z}
(\z\times\q)^i=-\ii \varepsilon^{imn}\frac{\partial}{\partial q^m}[q^n\delta(\q)],
\ee
which after integration by parts, where boundary terms trivially vanish,  immediately shows that
$\TT{\text{Diag. \ref{2nd_order_orbital}b}}=0$. This  implies 
\be
\text{Diag. \ref{2nd_order_orbital}b}=0.
\label{llllp}
\ee
We mention in passing that such a derivation of this result avoids singular
expressions that may be encountered after employment of (\ref{qwwww}).

{\bf Diagrams \ref{2nd_order_orbital}c and \ref{2nd_order_orbital}d}. These
diagrams, 
\begin{align}
&\text{Diag. \ref{2nd_order_orbital}c}=\limT \frac{\eo^2}{\Vol}
\intT\dd{x}\dd{y}\int\ddd{z}D_{\mu\nu}(x-y) \trr{S(y-z)\nabla_\z^i S(z-y)\gamma^\nu} 
\ous\gamma^\mu \us,
\label{4c}\\
&\text{Diag. \ref{2nd_order_orbital}d}=\limT\eo^2
\intT\dd{x}\dd{y}\int\d{z}
D_{\mu\nu}(x-y)\trr{S(x-z)\nabla_\z^iS(z-y)\gamma^\nu S(y-x)\gamma^\mu}, 
\label{4d}
\end{align}
do not contribute to fermionic orbital angular momentum 
because they are  $\sz$-independent--see identity (\ref{ubaru0}) 
and the discussion below (\ref{toot}).

The final IR-regularized result for fermionic  orbital angular momentum is
\be
\expval{J^i_\orb}{\Opr}{\lambda}=
\text{Diag. \ref{2nd_order_orbital}a}.
\label{z1234}
\ee

\subsection{Electromagnetic spin angular momentum}
\label{El_spin_sec}
We set $\chi=\spel$ in  (\ref{Jpert_chia}) and note that 
the matrix element, which we need to compute, factorizes into the product of
electromagnetic and fermionic matrix elements
\begin{subequations}
\begin{align}
&\la\0s|\T(\J^I_\spel)^i {\cal H}^{I}_\IN(x){\cal H}^{I}_\IN(y)|\0s\ra=
\eo^2{\cal A}^i_{\mu\nu}(x,y){\cal F}^{\mu\nu}(x,y),\\
\label{Aimunu}
&{\cal A}^i_{\mu\nu}(x,y)=\varepsilon^{imn}\int\d{z}\langle0|\T\N{F^I_{m0}(z) A^I_n(z)}  A^I_\mu(x)A^I_\nu(y)|0\rangle,\\
&{\cal F}^{\mu\nu}(x,y)=\la\0s|\T\N{\overline{\psi}_I(x)\gamma^\mu\psi_I(x)}\N{\overline{\psi}_I(y)\gamma^\nu\psi_I(y)}|\0s\ra.
\label{Fmunu}
\end{align}
\end{subequations}

Evaluation  of its  fermionic part was done in \cite{BDfield}, and we quote the final result  for completeness here 
\begin{subequations}
\be
{\cal F}^{\mu\nu}(x,y)={\cal F}_\sym^{\mu\nu}(x,y)+{\cal F}_\asym^{\mu\nu}(x,y),
\label{Fsymasym}
\ee
\ba
{\cal F}_\sym^{\mu\nu}(x,y)=& \frac{\ii}{(2\pi)^3}\int \dddd{p} 
\frac{p^\mu\eta^{\nu0}+p^\nu\eta^{\mu0}-p^0\eta^{\mu\nu}
+ \mo\eta^{\mu\nu}}{p^2-\mo^2+\izero} 
e^{\ii(f-p)\cdot (x-y)}
\\
+&2V \int  \dddd{p}   \dddd{q}
\frac{p^\mu q^\nu+p^\nu q^\mu-\eta^{\mu\nu}(p\cdot q-\mo^2)}{\B{p^2-\mo^2+\izero}
\B{q^2-\mo^2+\izero}} 
e^{\ii(p-q)\cdot(x-y)}
\\
+& (x\leftrightarrow y  \ \text{on all terms}),
\label{Fsym}
\ea
\be
{\cal F}^{\mu\nu}_\asym(x,y)= \frac{2\sz}{(2\pi)^3}\int \dddd{p}
\frac{
\varepsilon^{0\mu\nu3}\mo-\varepsilon^{\sigma\mu\nu3}p_\sigma
}{p^2-\mo^2+\izero} 
e^{\ii(f-p)\cdot(x-y)}
- (x\leftrightarrow y).
\ee
\label{ploik}%
\end{subequations}
The above splitting is based on symmetry (anti-symmetry) of 
${\cal F}^{\mu\nu}_\sym$  (${\cal F}^{\mu\nu}_\asym$) with respect to the transformation
$\mu\leftrightarrow\nu$. 
Another important difference between ${\cal F}^{\mu\nu}_\sym$  and ${\cal F}^{\mu\nu}_\asym$
is that the former is $\sz$-independent, and so it cannot contribute to the
final result due to reasons explained below (\ref{toot}). We will thus replace ${\cal F}^{\mu\nu}$ below by ${\cal
F}^{\mu\nu}_\asym$.

Electromagnetic matrix element (\ref{Aimunu}) is  easily obtained through Wick's theorem combined  
with the following identity
\be
\la0|\T \partial_\alpha A^I_\beta(x)A^I_\gamma(y)|0\ra=\frac{\partial}{\partial x^\alpha}D_{\beta\gamma}(x-y),
\label{bgt}
\ee
which can be shown with canonical commutation relations. It reads
\begin{subequations}
\be
\begin{aligned}
{\cal A}^i_{\mu\nu}(x,y)&=
\varepsilon^{imn}\int\d{z}\contraction{}{F}{^I_{m0}(z) }{A}
F^I_{m0}(z) A^I_\mu(x)
\contraction{}{A}{^I_\delta(z)}{A}A^I_n(z)A^I_\nu(y)+(\barexymunu)\\
&= \int   \dddd{p}\frac{dq^0}{2\pi}
\frac{a^i_{\mu\nu}(p,q^0)
e^{-\ii p\cdot x+\ii\tilde{q}\cdot y}}{(p^2-\lambda^2+\izero)(\tilde{q}^2-\lambda^2+\izero)}
,
\end{aligned}
\label{mat_spel}
\ee
\ba
a^i_{\mu\nu}(p,q_0)=&\ii\varepsilon^{imn}p_m\left[\eta_{0\nu}\eta_{n\mu}-\eta_{0\mu}\eta_{n\nu}+\xxii
\B{
\frac{q_0p_\mu\eta_{n\nu}}{p^2-\lambda^2/\xi+\izero}
-\frac{p_0\tilde{q}_\nu\eta_{n\mu}}{\tilde{q}^2-\lambda^2/\xi+\izero}
}\right]\\
+&\ii\varepsilon^{imn}
\eta_{m\mu}\eta_{n\nu}(p_0+q_0),
\label{uyuyuy}
\ea
\label{aimunu}%
\end{subequations}
where $\tilde{q}$ is defined in (\ref{tildeq}).

The IR-regularized expression for electromagnetic spin angular momentum of the electron can be then written
as 
\be
\expval{J^i_\spel}{\Opr}{\lambda}=-\frac{\eo^2}{2\Vol}\int\dd{x}\dd{y} 
{\cal A}^i_{\mu\nu}(x,y){\cal F}^{\mu\nu}_\asym(x,y).
\ee
After    simple algebra, we end up with a rather surprisingly compact formula
\be
\expval{J^i_\spel}{\Opr}{\lambda}
=-4\ii\eo^2\spinz\int\dddd{p}\frac{2(p^0-\mo)^2-(p_1)^2-(p_2)^2}{(p^2-\mo^2+\izero)
[(p-f)^2-\lambda^2+\izero]^2}.
\label{bareJel1}
\ee

\subsection{Electromagnetic orbital angular momentum}
\label{El_orb_sec}
We set $\chi=\orbel$ in  (\ref{Jpert_chia}) and again notice that the
resulting  matrix element, which   has to be computed,  factorizes into the product of
electromagnetic and fermionic matrix elements
\be
\la\0s|\T(\J^I_\orbel)^i {\cal H}^{I}_\IN(x){\cal H}^{I}_\IN(y)|\0s\ra=
\eo^2\BB{ 
{\cal B}^i_{\mu\nu}(x,y)
+{\cal C}^i_{\mu\nu}(x,y)
}
{\cal F}^{\mu\nu}(x,y),
\label{Bimunu}
\ee
where ${\cal B}^i_{\mu\nu}$ and ${\cal C}^i_{\mu\nu}$ will be defined below.

To compute the electromagnetic matrix element, equal to the expression in square brackets in (\ref{Bimunu}), 
we need to evaluate 
\be
\begin{aligned}
\int \d{z} &z^m\langle0|\T\N{\partial_\alpha A^I_\beta(z)\partial_\gamma A^I_\delta(z)}  A^I_\mu(x)A^I_\nu(y)|0\rangle\\
&=\int \d{z} z^m
\contraction{\partial_\alpha}{A}{^I_\beta(z) }{A}\partial_\alpha A^I_\beta(z) A^I_\mu(x)
\contraction{\partial_\gamma}{A}{^I_\delta(z)  }{A}\partial_\gamma
A^I_\delta(z)  A^I_\nu(y)+(\barexymunu)\\
&=-\int \d{z} z^m \dddd{p} \dddd{q}
\frac{p_\alpha d_{\beta\mu}(p)}{p^2-\lambda^2+\izero}\frac{q_\gamma d_{\delta\nu}(q)}{q^2-\lambda^2+\izero}
e^{-\ii p\cdot x +\ii q\cdot y +\ii (p-q)\cdot z} 
+
(\barexymunu)\\
&=
\int \dddd{p} \frac{dq^0}{2\pi}
 (\x+\y)^m\,\BB{ 
 {_{\alpha\beta\gamma\delta}\Box_{\mu\nu}}(p,\tilde{q})
 +{_{\alpha\beta\gamma\delta}\Box_{\nu\mu}}(\tilde{q},p)} 
 e^{-\ii p\cdot x+\ii \tilde{q}\cdot y}
 \\ 
&+
\int \dddd{p} \frac{dq^0}{2\pi} 
\BB{ 
{_{\alpha\beta\gamma\delta m}\cancel{\Box}_{\mu\nu}(p,\tilde{q})}
-
{_{\alpha\beta\gamma\delta m}\cancel{\Box}_{\nu\mu}(\tilde{q},p)}
}
e^{-\ii p\cdot x+\ii\tilde{q}\cdot y}
,
\end{aligned}
\label{ddAAAA}
\ee
where  contractions have been computed  as in (\ref{bgt}),   $\d{z}$ integration  has been done with 
\be
\int\ddd{z} 
z^m
e^{\ii (p-q)\cdot z} 
=\frac{\ii}{2}
\B{  \frac{\partial}{\partial p^m}  -  \frac{\partial}{\partial q^m}  }\delta(\p-\q),
\label{dkdq}
\ee
integration by parts has been employed, and 
\begin{align}
\label{box12a}
&{_{\alpha\beta\gamma\delta}\Box_{\mu\nu}}(p,q)=
-\frac{1}{2}
\frac{p_\alpha d_{\beta\mu}(p)}{p^2-\lambda^2+\izero} \frac{q_\gamma d_{\delta\nu}(q)}{q^2-\lambda^2+\izero},  \\
&{_{\alpha\beta\gamma\delta m}\cancel{\Box}_{\mu\nu}}(p,q)= 
\frac{\ii}{2}
\B{\frac{\partial}{\partial p^m}-\frac{\partial}{\partial q^m}}
\B{\frac{p_\alpha d_{\beta\mu}(p)}{p^2-\lambda^2+\izero}
\frac{q_\gamma d_{\delta\nu}(q)}{q^2-\lambda^2+\izero}}
\label{box12b}
\end{align}
have been introduced.
We mention in passing that there are no  boundary terms  from such  integration
by parts.

We obtain  by combining  (\ref{Jorbel}), (\ref{Bimunu}), and  (\ref{ddAAAA})
\begin{subequations}
\begin{align}
&{\cal B}^i_{\mu\nu}(x,y)= \int \dddd{p} \frac{dq^0}{2\pi} 
(\x+\y)^m\, b^i_{m\mu\nu}(p,\tilde{q})
e^{-\ii p\cdot x+\ii\tilde{q}\cdot y},\\
&b^i_{m\mu\nu}(p,q)=\varepsilon^{imn}\BB{ 
{_{j0nj}}\Box_{\mu\nu}(p,q)  - {_{0jnj}}\Box_{\mu\nu}(p,q)} + (\mu\leftrightarrow\nu,p\leftrightarrow q),
\end{align}
\label{BBmunu}%
\end{subequations}
\begin{subequations}
\begin{align}
&{\cal C}^i_{\mu\nu}(x,y)= \int \dddd{p} \frac{dq^0}{2\pi} 
c^i_{\mu\nu}(p,\tilde{q})
e^{-\ii p\cdot x+\ii\tilde{q}\cdot y},\\
&c^i_{\mu\nu}(p,q)=\varepsilon^{imn}\BB{
{_{j0njm}}\cancel{\Box}_{\mu\nu}(p,q) 
 -{_{0jnjm}}\cancel{\Box}_{\mu\nu}(p,q)} - (\mu\leftrightarrow\nu,p\leftrightarrow q).
\end{align}
\label{CCmunu}%
\end{subequations}

Proceeding similarly as in Sec. \ref{El_spin_sec}, we write the IR-regularized
expression  for electromagnetic orbital angular momentum of the electron as 
\be
\expval{J^i_\orbel}{\Opr}{\lambda} =-\frac{\eo^2}{2\Vol}\int\dd{x}\dd{y}  
\BB{{\cal B}^i_{\mu\nu}(x,y)+{\cal C}^i_{\mu\nu}(x,y)}
{\cal F}_\asym^{\mu\nu}(x,y),
\label{p1234p}
\ee
where the contribution of ${\cal B}^i_{\mu\nu}$ to (\ref{p1234p})  vanishes because 
it is proportional to the term that has the same structure as the right-hand
side of  (\ref{exp_vanish}). We get after simple algebra 
\begin{multline}
\expval{J^i_\orbel}{\Opr}{\lambda}=-\frac{2\eo^2\sz}{\Vol}\int\dd{x}\dd{y}\int\dddd{k}\dddd{p}\frac{dq^0}{(2\pi)^4}
\frac{\varepsilon^{0\mu\nu3}\mo-\varepsilon^{\sigma\mu\nu3}k_\sigma}{k^2-\mo^2+\izero}
 c^i_{\mu\nu}(p,\tilde{q})\\
\cdot
 e^{\ii(f-k-p)\cdot x+\ii(k+\tilde{q}-f)\cdot y}.
\label{qqeenn}
\end{multline}

Finally, with the help of
\be
c^i_{\mu\nu}(p,p)=\frac{\ii\varepsilon^{imn}p_m
(\eta_{0\mu}\eta_{n\nu}-\eta_{0\nu}\eta_{n\mu})}{(p^2-\lambda^2+\izero)^2}
\B{1-\frac{1}{2\xi}\frac{p^2-\lambda^2}{p^2-\lambda^2/\xi+\izero}},
\label{fpp}%
\ee
we obtain 
\begin{multline}
\expval{J^i_\orbel}{\Opr}{\lambda}
=-4\ii\eo^2\spinz\int\dddd{p}
\frac{(p_1)^2+(p_2)^2}{(p^2-\mo^2+\izero)[(p-f)^2-\lambda^2+\izero]^2}\\
\cdot
\BB{1-\frac{1}{2\xi} \frac{(p-f)^2-\lambda^2}{(p-f)^2-\lambda^2/\xi+\izero}     }.
\label{tytyty}
\end{multline}

\subsection{Gauge-fixing  angular momentum}
\label{G_fix_sec}
We set $\chi=\xi$ in  (\ref{Jpert_chia}) and note that the resulting
expression can be obtained by straightforward modifications of 
calculations from  Sec.  \ref{El_orb_sec}. Namely, 
  $\exval{J^i_\xi}{\Opr}$ is given by the right-hand side of (\ref{qqeenn}) with
$c^i_{\mu\nu}$ being replaced by $\tilde c^{\,i}_{\mu\nu}$, whose diagonal
components are given by 
\be
\tilde c^{\,i}_{\mu\nu}(p,p)=\frac{\ii\varepsilon^{imn}p_m}{
(p^2-\lambda^2+\izero)(p^2-\lambda^2/\xi+\izero)}\BB{
\frac{\eta_{0\mu}\eta_{n\nu}-\eta_{0\nu}\eta_{n\mu}}{2}
+\xxii
\frac{p_0
(p_\mu\eta_{n\nu}-p_\nu\eta_{n\mu})
}{p^2-\lambda^2/\xi+\izero}
}.
\ee
This leads to the following IR-regularized expression for gauge-fixing angular
momentum of the electron
\be
\expval{J^i_\xi}{\Opr}{\lambda}
=-2\ii\eo^2\spinz\int\dddd{p}
\frac{(p_1)^2+(p_2)^2}{(p^2-\mo^2+\izero)
[(p-f)^2-\lambda^2+\izero]
[(p-f)^2-\lambda^2/\xi+\izero]}.
\label{JxiEx}
\ee

\section{Pauli-Villars regularization}
\label{PVV_sec}
We will discuss here  implementation of  the Pauli-Villars regularization 
in our  calculations (see \cite{RayskiPR1949,PauliRMP1949} for early works
on this technique as well as   \cite{BogolubovBook,ColemanBook} for its
variations).
In its simplest  version, it is based on the following 
modifications of  either fermionic propagator
 (\ref{prop_fer})
\be
\frac{\gamma\cdot p+\mo}{p^2-\mo^2+\izero}\to
\frac{\gamma\cdot p+\mo}{p^2-\mo^2+\izero}- \frac{\gamma\cdot p+M}{p^2-\Lambda^2+\izero}, 
\label{repl_fer}
\ee
where $M=\mo,\Lambda$, or electromagnetic propagator (\ref{prop_el})
\be
\frac{d_{\mu\nu}(p)}{p^2-\lambda^2+\izero}\to
\frac{d_{\mu\nu}(p)}{p^2-\lambda^2+\izero}- \ltoL,
\label{repl_el}
\ee
where the  replacement  $\lambda\to\Lambda$ is also applied  to $d_{\mu\nu}(p)$,
which   depends on $\lambda$ too.
The parameter $\Lambda$ is supposed to be taken to infinity upon removal of
the regularization.    
We have implemented these three ad hoc replacements,  finding that none of them leads to
total angular momentum of the electron 
that is independent of $\xi$. Calculations leading to
such a conclusion can be
performed by  technically straightforward extensions of 
studies presented in this paper and so we will not linger over them.

Failure of these  popular yet somewhat  arbitrary   regularization attempts 
means that we need a systematic approach, imposing the 
Pauli-Villars regularization  consistently all across calculations. 
One may thus consider modifications of the Lagrangian density
(see  \cite{Schwartz,Gupta} for textbook introduction to this
technique).
Such a bottom-up approach introduces  ghost fields,
say $\tilde A^\mu$ and $\tilde\psi$,
through the replacement  
\be
\begin{aligned}
{\cal L}\to\tilde{\cal L}=&-\frac{1}{4}F_{\mu\nu}F^{\mu\nu}   
    -\frac{\xi}{2}\B{\partial_\mu A^\mu}^2
   +\frac{\mph^2}{2}A_\mu A^\mu
   +\overline{\psi}\B{\ii\gamma^\mu\partial_\mu-\mo}\psi\\
   &+\frac{1}{4}\tilde F_{\mu\nu}\tilde F^{\mu\nu}
    +\frac{\xi}{2}\B{\partial_\mu\tilde A^\mu}^2
   -\frac{\Lambda^2}{2}\tilde A_\mu\tilde A^\mu
   +\overline{\tilde\psi}\B{\ii\gamma^\mu\partial_\mu-\Lambda}\tilde\psi\\
&-\eo (\overline{\psi}\gamma^\mu\psi + \overline{\tilde\psi}\gamma^\mu\tilde\psi)(A_\mu+\tilde A_\mu).
\label{PVL}
\end{aligned}
\ee
This  leads to the interaction-picture  density of the interaction Hamiltonian
\be
\tilde{\cal H}^I_\IN
= \eo (\N{\overline{\psi}_I\gamma^\mu\psi_I} +\N{\overline{\tilde\psi}_I\gamma^\mu\tilde\psi_I})
(A^I_\mu+\tilde A^I_\mu),
\label{Htilde}
\ee
which  has to be used in imaginary time evolutions.
Such  evolutions in our studies start from the state 
\be
|\bullet\ra=|\0 s\ra\otimes|\tilde 0\ra,
\label{bullet}
\ee
where $|\tilde 0\ra$ contains no ghost particles.

As we discuss in Appendix \ref{Pauli_sec}, replacements 
\be
{\cal H}^I_\IN \to \tilde{\cal H}^I_\IN, \ |\0 s\ra \to|\bullet\ra
\label{repl}
\ee
performed on (\ref{fghj}) 
regularize only expectation values of $J^i_\sp$ and
$J^i_\orb$. They are equivalent to modification (\ref{repl_el})  
of the   electromagnetic propagator in calculations from  Secs.
\ref{Fer_spin_sec} and \ref{Fer_orb_sec}.
The problem now is that replacements (\ref{repl}), 
when imposed on  (\ref{fghj}), {\it do not} regularize expectation values of $J^i_\spel$, $J^i_\orbel$,
and $J^i_\xi$. 

To overcome this difficulty, we first introduce  ghost angular momentum
operators $\tilde J_\chi^i$, which are  obtained from $J_\chi^i$ by replacing all fields
with their ghost counterparts.  Next, we  consider 
\be
J^i-\tilde J^i=\sum_\chi (J^i_\chi- \tilde J^i_\chi), 
\label{JmJ}
\ee
where $\chi$ is given by (\ref{chi_list}).
The expectation value of the left-hand side of (\ref{JmJ}), upon removal of
the regularization, should  yield  total angular
momentum of the electron. It  should be so because  ghost angular momentum should  not contribute in
such a limit (there are no  ghost particles in the unperturbed state of the
system and  
the $\Lambda\to\infty$ limit suppresses addition of  such particles to the
perturbed state).  

The idea now is to  compute the expectation value of
\be
J^i_\chi-\tilde J^i_\chi
\label{jjj}
\ee
in the system described by modified Lagrangian density (\ref{PVL}), and to treat the resulting
expression, say $\expval{J^i_\chi}{\Opr}{\lambda\Lambda}$,
as both the IR- and UV-regularized  expectation value of the operator 
$J^i_\chi$.
According to remarks presented below  (\ref{JmJ}),
such a regularization procedure should  not affect the value of total angular
momentum of the electron, and so it may be considered  as a prospective 
solution to  regularization  challenges that we face.

To put such  a scheme to the test, we marry up (\ref{fghj}) with (\ref{repl}), and replace 
$J^i_\chi$ in the resulting formula by (\ref{jjj})  getting 
\be
\expval{J^i_\chi}{\Opr}{\lambda\Lambda}=\expval{J^i_\chi}{\Opr}{\lambda}-\expval{J^i_\chi}{\Opr}{\Lambda}
\label{lL}
\ee
for all angular momenta that we study (see Appendix \ref{Pauli_sec} for
derivation of this formula).
For $\chi=\sp,\orb$ this is exactly what one obtains through replacements
(\ref{repl}) imposed on (\ref{fghj}) because those angular momenta are linear in electromagnetic
propagators--see the comment below (\ref{repl}). For $\chi=\spel,\orbel,\xi$, (\ref{lL}) does not correspond to
any of  above-mentioned  modifications of propagators. For example,
(\ref{lL}) is not equivalent to (\ref{repl_el}) because expressions for 
those angular momenta are quadratic in electromagnetic propagators.
It is thus evident that  such a  ghost subtraction  technique  extends the
standard Pauli-Villars approach  based solely on modifications 
of Lagrangian density (\ref{PVL}). We find it quite reassuring that these two methods agree 
for fermionic spin and orbital angular momenta, where the standard approach
works.

All in all, (\ref{lL}) delivers   the consistent  Pauli-Villars regularization of all
angular momenta that we study. Such a procedure, when individual regularized angular momenta are added up,
leads to  the  $\xi$-independent value of total angular
momentum  of the electron (Sec. \ref{Regularized_sec}).
It is perhaps worth to stress  that the fact
that we work with arbitrary $\xi>0$ allows us for a rather stringent test 
of  reliability of the regularization procedure that we use.
Indeed, the requirement of gauge invariance, within the family of all covariant
gauges, eliminates a great deal of presumably sensible Pauli-Villars-like
regularizations.

\section{One-loop radiative corrections}
\label{Regularized_sec}
To compute one-loop radiative corrections, we will use
subtraction procedure (\ref{lL}) to impose ultraviolet (UV) regularization onto 
expressions (\ref{diagram2a_2}), (\ref{suma3}), (\ref{4a4a}), 
(\ref{bareJel1}), (\ref{tytyty}), and (\ref{JxiEx}).
This step is necessary because without it those expressions do not have definite
values.
To simplify such obtained formulae,  products of propagators' 
denominators will be joined with  the following identities
\begin{subequations}
\begin{align}
&\frac{1}{AB}=\int_0^1 ds\frac{1}{[sA + (1-s)B]^2},\\
&\frac{1}{AB^2}=\int_0^1 ds \frac{2(1-s)}{[s A + (1-s) B]^3},\\
&\frac{1}{A^2B^2}=\int_0^1 ds\,\frac{6(1-s)s}{[s A + (1-s)B]^4},\\
&\frac{1}{ABC}=\int_0^1ds\int_0^{1-s}du\frac{2}{[sA+uB+(1-s-u)C]^3},
\end{align}
\label{p3p3}%
\end{subequations}
the timelike component of the $4$-vector $p$ will be shifted to make  resulting  denominators 
$p^2$ dependent, Lorentz averaging of  numerators will be implemented through replacements $p_\mu
p_\nu\to\eta_{\mu\nu}p^2/4$, and finally  Wick rotation will be  performed followed by
straightforward evaluation of  resulting Euclidean integrals.
Such obtained expressions  will be  compactly written 
after introduction of the following functions
\begin{align}
&\Delta_\chi=(1-s)^2+s(\chi/\mo)^2,\\
&\tilde\Delta_\chi=(1-s-u)^2 + (s+u/\xi)(\chi/\mo)^2.
\end{align}
Above-mentioned calculations will be  done  under tacit assumptions that
these functions are greater than zero for $\chi=\lambda,\Lambda$.

\subsection{Fermionic spin angular momentum}
\label{Regularized_fer_spin}
We will apply here procedure (\ref{lL})  to individual
diagrams introducing  
\be
\LMPH{\text{Diag. X}}= \text{Diag. X}-\mphtoL
\label{iiooiioo}
\ee
as the Pauli-Villars-regularized version of 
 IR-regularized only $\text{Diag. X}$ from Sec. \ref{Fer_spin_sec}. 
Note that  limit (\ref{limitT}) is already taken in (\ref{iiooiioo}).

Following steps outlined around (\ref{p3p3}), we get
\begin{multline}
\LMPH{\text{Diag.  \ref{2nd_order_licznik}a}}=
 \frac{\eo^2\spinz}{8\pi^2}\int_0^1 ds (1-s)
\BB{
\ln\frac{\Delta_\Lambda}{\Delta_\lambda} + (1+s^2) \B{\frac{1}{\Delta_\Lambda} -
\frac{1}{\Delta_\lambda}}}+
\frac{\eo^2\spinz}{8\pi^2}\xxii\ln\frac{\Lambda}{\lambda}
\label{d1_s}
\end{multline}
and
\begin{multline}
\LMPH{\text{Diag. \ref{2nd_order_licznik}b}}+
\LMPH{\text{Diag. \ref{2nd_order_licznik}c}}+
\LMPH{\text{Diag. \ref{2nd_order_mianownik}a}} =  \\
\frac{\eo^2\spinz}{8\pi^2}\int_0^1 ds\left[
s\ln\frac{\Delta_\lambda}{\Delta_\Lambda}
+2(2-s)(1-s)s\B{\frac{1}{\Delta_\lambda} -\frac{1}{\Delta_\Lambda}}\right]
-\frac{\eo^2\spinz}{8\pi^2}\xxii\ln\frac{\Lambda}{\lambda}.
\label{d2new}
\end{multline}

Integrals in these equations can be 
analytically evaluated, but  resulting expressions are not compact. We list them 
in Appendix \ref{Integrals_app}. Among other things,  they can be used
for showing that  unless $\xi$ is fine-tuned, (\ref{d1_s}) and (\ref{d2new}) are 
logarithmically divergent in both IR and UV upon removal of the regularization. 
 For $\xi=\infty$, the Landau gauge, these expressions are still IR divergent
 but UV finite.   
 For $\xi=1/3$, the Fried-Yennie gauge, (\ref{d1_s}) and (\ref{d2new}) are 
IR finite but UV divergent. Both features are typical of covariant gauge calculations.

Next, we take  limits of $\mph\to0$ and $\Lambda\to\infty$
on the sum of  (\ref{Jpert1_prim}), (\ref{d1_s}), and (\ref{d2new}) getting 
\be
\label{earlyJspin}
\exval{J^i_\sp}{\Opr} =\spinz\B{1- \frac{\eo^2}{8\pi^2}}.
\ee
 Using  $\eo=e+O(e^3)$, this can be written as
\be
\label{alphaJspin}
\exval{J^i_\sp}{\Opr} =\spinz\B{1- \frac{\alpha}{2\pi}}+O(\alpha^2).
\ee
This one-loop result for  fermionic spin angular
momentum of the electron
agrees with earlier studies \cite{LiuPRD2015,JiPRD2016}.
Several  remarks are in order now.

To begin, our calculations show that (\ref{earlyJspin})   is $\xi$-independent, i.e.,
one and the same in the family of all covariant gauges.
This becomes apparent even before removal of the regularization due to  trivial cancellation of 
last terms in (\ref{d1_s}) and (\ref{d2new}) when the sum of all diagrams
is considered. We find it interesting that  $\xi$-dependence in these equations
takes such a   simple form despite the fact that $\xi$ shows up  in 
the  denominator 
of   electromagnetic propagator (\ref{prop_el}). Indeed, 
one would naturally expect that after  joining propagators' denominators 
through (\ref{p3p3}), $\xi$-dependence will be transferred to the $\Delta_\chi$-like
function appearing under  the integral over  the auxiliary  parameter   $s$. This is actually
what happens in  intermediate stages of calculations, but
then  unforeseen simplifications occur allowing for trivial
evaluation of $\xi$-dependent parts of (\ref{d1_s}) and (\ref{d2new}).

Next, we remark  that (\ref{d2new}) is equal to  $\spinz(Z_2-1)$, where $Z_2$ is the renormalization
constant of the Dirac field. One can easily verify this statement in the Feynman gauge by
looking at Sec. 7.1 of \cite{PS}, where $Z_2(\xi=1)$ is computed. 
In the general covariant gauge, one can repeat 
calculations from \cite{PS} with propagator (\ref{prop_el}). 
Such obtained expression  for $Z_2(\xi)$  is
complicated, but it can be easily numerically checked that it also supports the above remark.
Appearance of $Z_2$ in (\ref{d2new}) is expected. For example, 
 a quick look at 
 Figs. \ref{2nd_order_licznik}b, \ref{2nd_order_licznik}c, and 
 \ref{2nd_order_mianownik}a 
reveals that  diagrams depicted there are similar in structure to 
 the ones encountered during evaluation of $Z_2$ from the study of the electron propagator in the 
 QED vacuum state \cite{PS}. 
 Finally, we mention that 
 the  $\xi\neq1$ correction to $Z_2(\xi)$, which can be  extracted  from the last term in (\ref{d2new}),
appears  also in \cite{JohnsonPRL1959}, where  calculations are  Pauli-Villars-regularized in a slightly 
different way.\footnote{The
 difference  comes from the fact that our regularization 
is consistently implemented throughout calculations, whereas the one in  \cite{JohnsonPRL1959}
is done ``by hand''.}

\subsection{Other angular  momenta}
\label{Regularized_other_sec}
We will apply here regularization procedure (\ref{lL}) to angular momenta
studied in Sec. \ref{Other_sec}.
This results in the following set of equations
\begin{multline}
\expval{J^i_\orb}{\Opr}{\lambda\Lambda}=-4\ii\eo^2\spinz \int\dddd{p} 
\frac{(p_1)^2+(p_2)^2}{(p^2-\mo^2+\izero)^2}\\
\cdot\BB{\frac{1}{(p-f)^2-\lambda^2+\izero}
\B{1+\xxiidwa\frac{p^2-\mo^2}{(p-f)^2-\lambda^2/\xi+\izero}}-\ltoL},
\label{qqqw}
\end{multline}
\begin{multline}
\expval{J^i_\spel}{\Opr}{\lambda\Lambda}
=-4\ii\eo^2\spinz\int\dddd{p}\frac{2(p^0-\mo)^2-(p_1)^2-(p_2)^2}{p^2-\mo^2+\izero}
\\ \cdot\BB{\frac{1}{[(p-f)^2-\lambda^2+\izero]^2}-\ltoL},
\label{qaz}
\end{multline}
\begin{multline}
\expval{J^i_\orbel}{\Opr}{\lambda\Lambda}
=-4\ii\eo^2\spinz\int\dddd{p}
\frac{(p_1)^2+(p_2)^2}{p^2-\mo^2+\izero}\\
\cdot\BB{
\frac{1}{[(p-f)^2-\lambda^2+\izero]^2}
\B{1-\frac{1}{2\xi} \frac{(p-f)^2-\lambda^2}{(p-f)^2-\lambda^2/\xi+\izero}}-\ltoL},
\label{vvvvv}
\end{multline}
\begin{multline}
\expval{J^i_\xi}{\Opr}{\lambda\Lambda}
=-2\ii\eo^2\spinz\int\dddd{p}
\frac{(p_1)^2+(p_2)^2}{p^2-\mo^2+\izero}
\\ \cdot\BB{
\frac{1}{[(p-f)^2-\lambda^2+\izero][(p-f)^2-\lambda^2/\xi+\izero]}-\ltoL
}.
\label{JxiExPV}
\end{multline}

Even without  evaluating  these expressions, 
one can  notice that their sum is $\xi$-independent, which is something
that we have anticipated in Sec. \ref{PVV_sec}. A bit surprising now is that
 $\expval{J^i_\spel}{\Opr}{\lambda\Lambda}$ and 
$\expval{J^i_\orb+J^i_\orbel+J^i_\xi}{\Opr}{\lambda\Lambda}$ are separately  $\xi$-independent.
Such an observation, however, is formal because we will shortly see that both
quantities are actually infinite upon removal of the regularization.

Following the procedure outlined at the beginning of Sec. \ref{Regularized_sec}, we get
\begin{align}
&\expval{J^i_\orb}{\Opr}{\lambda\Lambda}
=-\spinz\frac{\eo^2}{4\pi^2}\int_0^1ds (1-s)\ln\frac{\Delta_\Lambda}{\Delta_\lambda}
-\spinz\frac{\eo^2}{8\pi^2}\xxii\int_0^1ds\int_0^{1-s}du
\ln\frac{\tilde\Delta_\Lambda}{\tilde\Delta_\lambda},
\\
&\expval{J^i_\spel}{\Opr}{\lambda\Lambda}
=\spinz\frac{\eo^2}{2\pi^2}
\int_0^1ds\, s \BB{ \ln\frac{\Delta_\Lambda}{\Delta_\lambda}
-(1-s)^2\B{\frac{1}{\Delta_\lambda}  -\frac{1}{\Delta_\Lambda}  } },
\\
&\expval{J^i_\orbel}{\Opr}{\lambda\Lambda}
=-\spinz\frac{\eo^2}{4\pi^2}\int_0^1ds\, s  \ln\frac{\Delta_\Lambda}{\Delta_\lambda}
+\spinz\frac{\eo^2}{8\pi^2}\frac{1}{\xi}\int_0^1ds\int_0^{1-s}du
\ln\frac{\tilde\Delta_\Lambda}{\tilde\Delta_\lambda},
\\
&\expval{J^i_\xi}{\Opr}{\lambda\Lambda}
=-\spinz\frac{\eo^2}{8\pi^2}\int_0^1ds\int_0^{1-s}du
\ln\frac{\tilde\Delta_\Lambda}{\tilde\Delta_\lambda}.
\end{align}

This can be further simplified if we remove the IR regularization. With some 
extra effort, we get the following results exhibiting rather non-trivial
$\xi$-dependence
\begin{align}
&\limlzero\expval{J^i_\orb}{\Opr}{\lambda\Lambda}
\simeq\spinz\frac{\eo^2}{8\pi^2}\B{-\frac{1+\xi}{\xi}\ln\frac{\Lambda}{\mo}+\frac{5}{4}
-\frac{3}{4\xi}+\frac{\ln\xi}{2\xi}},\\
&\limlzero\expval{J^i_\spel}{\Opr}{\lambda\Lambda}\simeq\spinz\frac{\eo^2}{2\pi^2}
\B{\ln\frac{\Lambda}{\mo}+\frac{3}{4}},\\
\label{JorbelAss}
&\limlzero\expval{J^i_\orbel}{\Opr}{\lambda\Lambda}
\simeq\spinz\frac{\eo^2}{8\pi^2}\B{\frac{1-2\xi}{\xi}\ln\frac{\Lambda}{\mo}-\frac{5}{2}
+\frac{3}{4\xi}-\frac{\ln\xi}{2\xi(1-\xi)}},\\
\label{JxiAss}
&\limlzero\expval{J^i_\xi}{\Opr}{\lambda\Lambda}\simeq\spinz\frac{\eo^2}{8\pi^2}\B{-\ln\frac{\Lambda}{\mo}-\frac{3}{4}
+\frac{\ln\xi}{2(1-\xi)}},
\end{align}
where $\simeq$ means that we omit terms that vanish in the
limit of $\Lambda\to\infty$.
Note that all these expressions are well-defined for any $\xi>0$. 
Among other things, they allow us to conclude that  upon removal of the regularization 
\be
\exval{J^i_\orb+J^i_\spel+J^i_\orbel+J^i_\xi}{\Opr}=\spinz\frac{\eo^2}{8\pi^2}.
\label{qqqmmm}
\ee

Combining (\ref{qqqmmm})  with (\ref{earlyJspin}), we see that in our one-loop calculations 
the expectation value of total angular momentum operator (\ref{totalJ})  is 
given by (\ref{toot}), which  can be seen as a self-consistency check of our
studies. 

\section{Discussion}
\label{Discussion_sec}

We have teamed the bare perturbative expansion with the imaginary time evolution
technique  to  study radiative corrections to different components of 
angular momentum of the electron. Our calculations have been done in the
general covariant gauge. The results that  we have obtained can be summarized as follows.

First, we have carefully discussed implementation of  imaginary time
evolutions developing a rigorous analytical procedure  taking care of singularities 
that may appear in the course of calculations. 
Such  evolutions are routinely  used for generation of ground states,
which are then used for computation of expectation
values of products of field operators
in interacting 
quantum field theories. Results that we present on this matter are 
 missed in standard textbooks on quantum field theory, where enforcement of the imaginary
time limit  is trivialized to  steps outlined 
between (\ref{dx0sin}) and  (\ref{deltac}). 
On the one hand, our calculations show how disastrous such an oversimplification  is when bare
perturbation theory is employed for evaluation of  self-energy-type diagrams. 
On the other hand, they provide a general framework
that can be readily deployed in computations of other expectation values 
in quantum field theories. This can be useful   for either resolving 
possible issues with ``simplified'' handling of the imaginary time limit or
 for rigorous checking whether such a procedure is justified. 
These  remarks are comprehensively  illustrated by our studies in Sec. \ref{Fer_spin_sec},
where  computations of some diagrams have been only possible after   sophisticated enforcement of
the imaginary time limit.

Second, we have computed fermionic spin and orbital, electromagnetic spin and
orbital, and gauge-fixing angular momenta of the electron. Out of  these five quantities, 
only  fermionic spin angular momentum is gauge invariant, and so it can be
conclusively  compared to  earlier studies, which were done
in the light-cone gauge \cite{LiuPRD2015,JiPRD2016}. It agrees with these works showing
 equivalence of the light-cone and  
general covariant gauge calculations. While such an agreement is expected on
general grounds, it is perhaps worth to mention that 
the issue of gauge independence is still  quite non-trivial (Sec. 2.5.2 of
\cite{LeaderPhysRep2014}).
More importantly,  technical comparision between  calculations in these completely different
gauges should be interesting and our detailed discussion should  facilitate it.

Third, the remaining four angular momenta are gauge non-invariant. Out of them,
gauge-fixing angular momentum is specific to  covariant 
gauge studies and it is instructive to take a closer look at it. It is so
because its presence turns out to be of key importance 
to assigning  spin one-half to the electron in  covariantly quantized
electrodynamics.
Indeed,  (\ref{toot}) would not hold without it
even in  the  $\xi\to\infty$ limit, 
where the Lorentz gauge is most transparently enforced (Sec. 15.5 of \cite{WeinbergII}).
This is interesting because  $J^i_\xi$ can be seen
as a physically meaningless artifact of the quantization procedure and so the
question arises why it non-trivially contributes to the physically meaningful quantity
such as electron's spin.
We expect  that  resolution of this puzzle is the following. The
gauge-fixing term in  Lagrangian density  (\ref{LL})
not only generates  gauge-fixing angular momentum, but it also affects 
the electromagnetic propagator. The latter impacts computations of expectation
values of gauge non-invariant angular momentum operators. As a result,  
those expectation values get implicitly modified by the presence of the gauge-fixing
term and this modification is explicitely cancelled in (\ref{toot}) by gauge-fixing
angular momentum, so that it has no effect on electron's spin.

Fourth, we have developed a variant of the Pauli-Villars regularization by
requiring that  total angular momentum of the electron should be one and the
same in
the family of all covariant gauges. This  obvious condition is
violated by  the simplest versions of the Pauli-Villars regularization.
In our scheme, one subtracts  from the observable of interest
its ghost operator counterpart, and then calculates the expectation value of 
such obtained operator  through imaginary time evolution.
The latter is consistently implemented by the standard 
addition of ghost fields to  the Lagrangian density.
The net effect of this procedure is very  simple for  observables that we study  (\ref{lL}). 
We believe that it 
would be   interesting to put   this approach to the test in other problems as
well.

Finally, to place   result  (\ref{alphaJspin}) in a wider context, we mention that
 only one more finite  gauge invariant individual component of total angular momentum of the
electron was identified  so far. Namely, electromagnetic angular
momentum \cite{BDfield}
\be
\LARA{\int \d{z} \BB{\z\times(\E\times\Bold)}^i}_{\Opr}
= -\spinz\frac{\alpha}{2\pi}+O(\alpha^2),
\label{Jfield}
\ee
where $\E$ and $\Bold$ are electric and magnetic field operators.\footnote{Such a result was    obtained with 
ad hoc regularization attempts (\ref{repl_fer}) and (\ref{repl_el}) 
explored in \cite{BDfield}. It can be also obtained 
with the ghost subtraction technique discussed in  Sec. \ref{PVV_sec} and Appendix \ref{Pauli_sec} of this paper.
Its indifference to  details of the Pauli-Villars regularization scheme
presumably comes 
from favorable convergence properties of the expression that is regularized
during evaluation of (\ref{Jfield}).}
Gauge invariance and finiteness of (\ref{alphaJspin}) and (\ref{Jfield}) should make them 
especially interesting from the experimental point of view.
Given the fact that
various angular momenta, contributing  to nucleons' spin,  have been extensively experimentally studied
\cite{Deur2019}, we are  hopeful that 
such  quantities   can be  also measured. The remaining open question is how
this can be achieved.

\noindent{\bf Acknowledgements}\\ 
\noindent 
I would like to thank Aneta for being a wonderful sounding board during
all these studies.
Diagrams in this work have been   done in JaxoDraw \cite{JaxoDraw2}.
This work has been  supported by the Polish National Science Centre (NCN) grant DEC-2016/23/B/ST3/01152.

\appendix

\section{Conventions and all that}
\label{Conventions_sec}
We use the Minkowski metric $\eta=\text{diag}(+---)$ and choose $\varepsilon^{0123}=+1=\varepsilon^{123}$.
Greek and Latin indices  take values $0,1,2,3$ and   $1,2,3$, respectively,
when they refer to components of $4$- and $3$-vectors. 
We  use the Einstein summation convention.
3-vectors are written  in bold,  e.g. $x=(x^\mu)=(x^0,\x)$.
Electron's bare and physical charges are both negative.

We introduce  
\be
\exval{\cdots}{\Psi}=\frac{\la\Psi|\cdots|\Psi\ra}{\la\Psi|\Psi\ra}, \ 
\om{q}=|\q|,  \ \vareps{q}=\sqrt{\mo^2+\om{q}^2},
\label{expectation}
\ee
and  write the interaction-picture Dirac field operator as
\begin{subequations}
\begin{align}
\label{DiracI}
&\psi_I(x)=\int\frac{\d{p}}{(2\pi)^{3/2}} \sqrt{\frac{\mo}{\vareps{p}}}
\sum_s\BB{a_{\p s}u(\p,s)e^{-\ii p\cdot x} + b^\dag_{\p s} v(\p,s)e^{\ii p\cdot x}  },  \ 
 (p^\mu)=(\vareps{p},\p),\\
&\{a_{\p s},a^\dag_{\q r}\}=\{b_{\p s},b^\dag_{\q r}\}=\delta_{sr}\delta(\p-\q), 
\end{align}
\end{subequations}
where $a_{\p s}$ annihilates the electron and 
$b_{\p s}$ annihilates  the positron (both of momentum $\p$ and the spin state  $s$). 
All other anticommutators involving those  operators are equal to zero.
We choose bispinors $u(\p,s)$ and $v(\p,s)$, in the standard representation of $\gamma$ matrices
that we use, so that 
\begin{subequations}
\begin{align}
\label{ups}
&u(\p,s)=
\frac{1}{\sqrt{2\mo(\vareps{p}+\mo)}} 
\binom{(\vareps{p}+\mo)\phi^s}{\p\cdot\sig\phi^s},  \ v(\p,s)= \frac{1}{\sqrt{2\mo(\vareps{p}-\mo)}}
\binom{(\vareps{p}-\mo)\phi^s}{\p\cdot\sig\phi^s},\\
&\phi^s=\binom{1}{0},\binom{0}{1}.
\end{align}
\end{subequations}

We define contractions of  $\psi_I$  on zero-momentum external lines as
\be
\contraction{}{\psi}{_I(x)|}{\0 s}
\psi_I(x)|\0 s\ra=\frac{\us}{(2\pi)^{3/2}}e^{-\ii f\cdot x},  \ 
\contraction{\la}{\0 s}{|}{\overline{\psi}}
\la\0 s|\overline{\psi}_I(x)=\frac{\ous}{(2\pi)^{3/2}}
e^{\ii f\cdot x},  \ \us =u(\0,s),
\label{ext_contractions}
\ee
where $|\0 s\rangle$ and $f$ are given by (\ref{0s}) and (\ref{f0}),
respectively.
The $\us$  bispinors are eigenstates of the $z$-component of the one-particle fermionic spin angular momentum 
operator  
\be
\frac{1}{2}\Sigma^3 \us = \sz \us, 
  \ \us=\left(
\begin{array}{l}
1\\0\\0\\0
\end{array}
\right) \ \text{for} \ \sz=+1/2,  \ 
\us=\left(
\begin{array}{l}
0\\1\\0\\0
\end{array}
\right) \ \text{for} \ \sz=-1/2. 
\label{u}
\ee

Finally, we mention that there is no summation over $s$ in matrix elements $\ous\cdots\us$.

\section{Bispinor matrix elements}
\label{Matrix_sec}
Results presented below are obtained in the standard (Dirac) representation of $\gamma$ matrices.
It is then  a simple exercise to show that the same results are  obtained in all 
representations  unitarily  similar to the standard one (Weil, Majorana, etc.).
This statement is equivalent
to saying that they  are invariant under   
$\gamma^\mu\to U\gamma^\mu U^\dag$ and $\us\to U\us$ transformations, where  
$U$ is an arbitrary   unitary matrix of dimension four (see
\cite{PalArxiv2007,ArminjonBraz2008} 
for the discussion of  representation-independence of various results associated
with the Dirac equation).

The following expressions are used in our computations
\begin{gather}
\ous\gamma^\mu(\gamma\cdot p+\mo)\gamma_\mu \us=4\mo-2p^0,
\label{ubaru2}\\
\ous\gamma\cdot k(\gamma\cdot p+\mo)\gamma\cdot k\us=2k^0k\cdot p+k^2(\mo-p^0), 
\label{xiubaru4}\\
\ous\Gamma^i(\gamma^0q^0+\mo)\gamma^\mu(\gamma\cdot p+\mo)\gamma_\mu \us=
\spinz(\mo+q^0)\ous\gamma^\mu(\gamma\cdot p+\mo)\gamma_\mu \us,
\label{ubaru4}\\
\ous\Gamma^i(\gamma^0 q^0+\mo)\gamma\cdot k (\gamma\cdot
p+\mo)\gamma\cdot k\us=\spinz(\mo+q^0)\ous\gamma\cdot k(\gamma\cdot p+\mo)\gamma\cdot k\us,
\label{xiubaru2}\\
\ous\gamma^\mu(\gamma\cdot p+\mo)\Gamma^i(\gamma\cdot p+\mo)\gamma_\mu
\us=2\sz\BB{\delta^{i3}(p^2+\mo^2)+2p_i p_3},  
\label{ubaru6}\\
\ous\gamma\cdot(f- p)(\gamma\cdot p+\mo)\Gamma^i (\gamma\cdot p+\mo)\gamma\cdot(f-p)\us=
\spinz (p^2-\mo^2)^2,
\label{xiubaru6}\\
\label{ubaru0}
\ous\gamma^\mu \us=\eta^{\mu0},\\
\label{ubaru12}
\varepsilon^{imn}p^n\ous\gamma^\mu\{\gamma^m\gamma^0,\gamma\cdot p+\mo\}\gamma_\mu\us=-8\ii\sz(\delta^{i3}\om{p}^2-p_ip_3),\\
\label{ubaru12next}
\varepsilon^{imn}p^n\ous\gamma\cdot(f-p)\{\gamma^m\gamma^0,\gamma\cdot p+\mo\}\gamma\cdot(f-p)\us=-4\ii\sz(\delta^{i3}\om{p}^2-p_ip_3)(p^2-\mo^2).
\end{gather}
We mention in passing that we simplify matrix elements (\ref{ubaru6}), (\ref{ubaru12}), and (\ref{ubaru12next}) 
under  integral signs by the replacement
$p_ip_3\to\delta^{i3}(p_3)^2$.

It is interesting to note that $\sz$-dependence, in all expectation values that we study,
comes from expressions  that critically depend on
the four-dimensional Levi-Civita symbol, whose extension to a $d\neq4$ dimensional 
space-time, used in the dimensional regularization, is problematic (see e.g.
Appendix B.2 of \cite{DreinerPhysRep2010}
and references therein). This can  be proved by combining (\ref{ubaru0}) and the
following easy-to-verify identities 
\begin{gather}
\ous\gamma^\mu\gamma^\nu\us=\eta^{\mu\nu}-2\ii\sz\varepsilon^{0\mu\nu3},
\label{2gamma}\\
\ous\gamma^\mu\gamma^\sigma\gamma^\nu\us=
\eta^{\mu\sigma}\eta^{\nu0}+\eta^{\sigma\nu}\eta^{\mu0}-\eta^{\mu\nu}\eta^{\sigma0}-2\ii\sz\varepsilon^{\mu\sigma\nu3},
\label{3gamma}\\
\ous\gamma^0\gamma^1\gamma^2\gamma^3\us=0
\label{gamma5}
\end{gather}
 with the observation  that 
 any product of $\gamma$ matrices can be always reduced to the single term
 containing at most  four  $\gamma$ matrices, whose  indices 
 are distinct.

\section{Implementation of imaginary time evolutions}
\label{Implementation_sec}
In the following, we  work out  integrals that are   necessary for  implementation of imaginary time evolutions. 
While doing so,  we will frequently use  the Sochocki-Plemelj formula
\be
\dashint dx \frac{f(x)}{x-x_0} = \int dx\BB{\pm\ii\pi\delta(x-x_0)+ \frac{1}{x-x_0\pm \izero}}f(x),
\label{SP}
\ee
where $\dashint$ stands for the Cauchy principal value. 
Several things have to be kept in mind in the following discussion.

First,  as we have mentioned in Sec. \ref{Basics_sec}, $T$ will be greater
than zero during
evaluation of integrals and then the limit $\barelimT$ will be taken.

Second, we will use   below the function 
\be
G(k^0,p^0,\dots),
\label{Gkp}
\ee
which will be assumed to have poles at 
\be
k^0=\pm\sqrt{\om{p}^2+M^2}\mp\izero, \ p^0=\pm\sqrt{\om{p}^2+M'^2}\mp\izero, 
\label{poles}
\ee
etc. Masses $M$, $M'$, etc. will
be greater than  zero. 
In other words,  poles of (\ref{Gkp}) will
come from propagators' denominators: $(k^0)^2-\om{p}^2-M^2+\izero$, $(p^0)^2-\om{p}^2-{M'}^2+\izero$,
etc. 

Third,  as (\ref{Gkp}) will vanish for large arguments in our studies, there will be no problems with
convergence of  contour integrals that we will discuss. 

{\bf Type $\Rzymskie{1}$ integrals}. The integrals of interest here are given
by the formula
\be
\chi_\Rzymskie{1}
=  \limT \int dp^0 dk^0 G(k^0,p^0) \frac{\sin^2[T(k^0+p^0-\mo)]}{(k^0+p^0-\mo)^2},
\label{chistart}
\ee
where  poles of the function $G$ are characterized by
$M>0$ and $M'=\mo$. Such integrals  appear in studies of Diag.
\ref{2nd_order_mianownik}a, where 
$M$ is  greater than zero due to the IR regularization
provided by either the
photon mass term or the  ghost photon mass term in Pauli-Villars-regularized calculations.

We rewrite (\ref{chistart})  as
\be
\begin{aligned}
\chi_\Rzymskie{1} 
=&\frac{1}{4} \limT \dashint dp^0  dk^0 G(k^0,p^0) \frac{1-e^{2\ii T(k^0+p^0-\mo)}}{k^0+p^0-\mo} \frac{1}{k^0+p^0-\mo}\\
+&\frac{1}{4} \limT \dashint dp^0  dk^0 G(k^0,p^0) \frac{1-e^{-2\ii T(k^0+p^0-\mo)}}{k^0+p^0-\mo} \frac{1}{k^0+p^0-\mo}.
\end{aligned}
\ee

Using now (\ref{SP}), we arrive at
\begin{subequations}
\begin{align}
\chi_\Rzymskie{1} 
&=\pi \int dp^0 G(\mo-p^0,p^0) \limT  T
\label{chii_a} \\
&+\frac{1}{4} \limT\int dp^0 dk^0 G(k^0,p^0)\frac{1-e^{2\ii T(k^0+p^0-\mo)}}{k^0+p^0-\mo}\frac{1}{k^0+p^0-\mo+\izero} \label{chii_b}\\
&+\frac{1}{4}\limT\int dp^0 dk^0 G(k^0,p^0)\frac{1-e^{-2\ii T(k^0+p^0-\mo)}}{k^0+p^0-\mo} \frac{1}{k^0+p^0-\mo-\izero}.\label{chii_c}
\end{align}
\label{chii}%
\end{subequations}

Suppose now that we evaluate integrals (\ref{chii_b}) and (\ref{chii_c})  on   semicircular contours
in  upper and lower half-planes of complex $k^0$  and $p^0$,
respectively.  This  turns exponential terms in (\ref{chii_b}) and (\ref{chii_c})
into 
\begin{subequations}
\begin{align}
&e^{\pm2\ii T(k^0+p^0-\mo)} \xrightarrow[\text{integrations}]{\text{contour}} e^{-2\ii T \gamma_\pm},
	\\
&\gamma_\pm=\sqrt{\om{p}^2+M^2}+\sqrt{\om{p}^2+{M'}^2}\pm\mo.
\end{align}
\label{exp0}%
\end{subequations}

Next, we note that $\gamma_\pm>0$ for $M$ and  $M'$ specified below
(\ref{chistart}).
Therefore, when we take the limit $\barelimT$, 
 exponential terms can be dropped from (\ref{chii_b}) and (\ref{chii_c}) if we properly shift 
 poles of $1/(k^0+p^0-\mo)$, which amounts to
\ba
\chi_\Rzymskie{1} 
=&\pi \int dp^0 G(\mo-p^0,p^0) \limT  T \\
+&\frac{1}{4} \int dp^0 dk^0 
\BB{\frac{G(k^0,p^0)}{(k^0+p^0-\mo+\izero)^2}+ \frac{G(k^0,p^0)}{(k^0+p^0-\mo-\izero)^2}}.
\label{chi}
\ea

{\bf Type $\Rzymskie{2}$ integrals}. Next, we introduce 
\be
\tilde G(k^0,p^0,q^0)=\frac{G(k^0,p^0)}{q^0-\mo+\izero},
\ee
where $G(k^0,p^0)$ is the same as in $\chi_\Rzymskie{1}$, 
and consider 
\be
\chi_\Rzymskie{2}
=  \limT \int dp^0 dk^0 dq^0 \tilde G(k^0,p^0,q^0) 
\frac{\sin[T(k^0+p^0-q^0)]}{k^0+p^0-q^0} \frac{\sin[T(k^0+p^0-\mo)]}{k^0+p^0-\mo}.
\label{uuu1234}
\ee
Integrals of such a form  appear in studies of Diags.
\ref{2nd_order_licznik}b and \ref{2nd_order_licznik}c.

We rewrite (\ref{uuu1234}) as
\ba
\chi_\Rzymskie{2}
= &\frac{1}{4} \limT \dashint dp^0 dk^0 dq^0 \tilde G(k^0,p^0,q^0)
e^{-\ii T(q^0-\mo)} 
\frac{1-e^{2\ii T(k^0+p^0-\mo)}}{k^0+p^0-\mo} \frac{1}{k^0+p^0-q^0}\\
+
&\frac{1}{4} \limT \dashint dp^0 dk^0  dq^0 \tilde G(k^0,p^0,q^0) 
e^{\ii T(q^0-\mo)} \frac{1-e^{-2\ii T(k^0+p^0-\mo)}}{k^0+p^0-\mo} \frac{1}{k^0+p^0-q^0}.
\ea

Employing  (\ref{SP}), we obtain
\ba
\chi_\Rzymskie{2}
= &\frac{1}{4} \limT \int dp^0 dk^0 dq^0 \tilde G(k^0,p^0,q^0)
e^{-\ii T(q^0-\mo)} 
\frac{1-e^{2\ii T(k^0+p^0-\mo)}}{k^0+p^0-\mo} \frac{1}{k^0+p^0-q^0+\izero}\\
+
&\frac{1}{4} \limT \int dp^0 dk^0 dq^0   \tilde G(k^0,p^0,q^0) 
e^{\ii T(q^0-\mo)} \frac{1-e^{-2\ii T(k^0+p^0-\mo)}}{k^0+p^0-\mo} \frac{1}{k^0+p^0-q^0+\izero}.
\label{2chii1}
\ea

Doing the first (second)  integral over $q^0$
on  the lower (upper) semicircular contour 
of the complex $q^0$ half-plane, joining  integrals, rearranging  terms, and then splitting them again we arrive at
\ba
\chi_\Rzymskie{2}
=&-\frac{\ii\pi}{2} \limT \dashint dp^0  dk^0 G(k^0,p^0) 
\frac{e^{\ii T(k^0+p^0-\mo)}-e^{2\ii T(k^0+p^0-\mo)}}{k^0+p^0-\mo}
\frac{1}{k^0+p^0-\mo}\\
 &-\frac{\ii\pi}{2} \limT \dashint dp^0  dk^0  G(k^0,p^0) 
\frac{1-e^{-\ii T(k^0+p^0-\mo)}}{k^0+p^0-\mo}
\frac{1}{k^0+p^0-\mo}.
\label{2chii2}
\ea

Using again (\ref{SP}), we obtain 
\be
\begin{aligned}
\chi_\Rzymskie{2}
&=-\ii\pi^2\int dp^0   G(\mo-p^0,p^0) \limT T\\
&-\frac{\ii\pi}{2} \limT \int dp^0  dk^0  G(k^0,p^0)\frac{e^{\ii T(k^0+p^0-\mo)}-e^{2\ii T(k^0+p^0-\mo)}}{k^0+p^0-\mo}
\frac{1}{k^0+p^0-\mo+\izero}\\
&-\frac{\ii\pi}{2} \limT \int dp^0  dk^0 G(k^0,p^0)\frac{1-e^{-\ii T (k^0+p^0-\mo)}}{k^0+p^0-\mo}
\frac{1}{k^0+p^0-\mo-\izero}.
\end{aligned}
\label{2chii3}%
\ee

Repeating now  steps around  (\ref{exp0}), we note that exponential terms in (\ref{2chii3})  
vanish upon taking the limit, which after proper shifting of the pole of
$1/(k^0+p^0-\mo)$ leaves us with
\be
\chi_\Rzymskie{2}
=-\ii\pi^2\int dp^0   G(\mo-p^0,p^0) \limT T
-\frac{\ii\pi}{2}  \int dp^0  dk^0 \frac{G(k^0,p^0)}{(k^0+p^0-\mo-\izero)^2}.
\label{2chii}%
\ee

{\bf Type $\Rzymskie{3}$ integrals}. 
Now, we consider 
\be
\chi_\Rzymskie{3}
=\limT\int dp^0 dk^0 dq^0 G(k^0,p^0,q^0)
\frac{\sin[T(k^0+p^0-\mo)]}{k^0+p^0-\mo} \frac{\sin[T(k^0+q^0-\mo)]}{k^0+q^0-\mo},
\label{chiii1}
\ee
where poles of $G(k^0,p^0,q^0)$ are parameterized  by
$M>0$ and  $M'=M''=\mo$ in expressions for  Diags. \ref{2nd_order_licznik}a and \ref{2nd_order_orbital}a.
During evaluation of electromagnetic spin, electromagnetic orbital, and
gauge-fixing angular momenta, they are given  by $M=\mo$, and $M',M''>0$.

We rewrite (\ref{chiii1})  as
\ba
\chi_\Rzymskie{3}
= 
&\frac{1}{2\ii}\limT\dashint dp^0 dk^0 dq^0 G(k^0,p^0,q^0) 
\frac{\sin[T(k^0+p^0-\mo)]}{k^0+p^0-\mo} \frac{e^{\ii T(k^0+q^0-\mo)}}{k^0+q^0-\mo}\\
-
&\frac{1}{2\ii}\limT\dashint dp^0 dk^0 dq^0 G(k^0,p^0,q^0)
\frac{\sin[T(k^0+p^0-\mo)]}{k^0+p^0-\mo} \frac{e^{-\ii T(k^0+q^0-\mo)}}{k^0+q^0-\mo},
\ea
which after using (\ref{SP}) leads to 
\ba
\chi_\Rzymskie{3}
= 
&\pi\limT\int dp^0 dq^0 G(\mo-q^0,p^0,q^0)
\frac{\sin[T(p^0-q^0)]}{p^0-q^0}\\
+&\frac{1}{2\ii}\limT\int dp^0 dk^0 dq^0 G(k^0,p^0,q^0)
\frac{\sin[T(k^0+p^0-\mo)]}{k^0+p^0-\mo} \\
&\pushright{\cdot\BB{\frac{e^{\ii T(k^0+q^0-\mo)}}{k^0+q^0-\mo+\izero}
- \frac{e^{-\ii T(k^0+q^0-\mo)}}{k^0+q^0-\mo-\izero}}}.
\ea

After splitting integrals over sinuses into  Cauchy principal value integrals and then one
more employment of (\ref{SP}), we obtain  
\begin{subequations}
\begin{align}
\chi_\Rzymskie{3}
&= 
\pi^2\int dp^0 G(\mo-p^0,p^0,p^0)\\
\label{bb}
+&\frac{\pi}{2\ii}\limT\int dp^0 dq^0 \BB{G(\mo-q^0,p^0,q^0)+G(\mo-p^0,p^0,q^0)}
\frac{e^{\ii T(p^0-q^0)}}{p^0-q^0+\izero}\\
+&\frac{\pi}{2\ii}\limT\int dp^0 dq^0 \BB{G(\mo-q^0,p^0,q^0)+G(\mo-p^0,p^0,q^0)}
\frac{e^{\ii T(q^0-p^0)}}{q^0-p^0+\izero}\\
-&\frac{1}{4}\limT\int dp^0 dk^0 dq^0 G(k^0,p^0,q^0) 
\frac{e^{\ii T(2k^0+p^0+q^0-2\mo)}}{(k^0+p^0-\mo+\izero)(k^0+q^0-\mo+\izero)}\\
\label{ee}
-&\frac{1}{4}\limT\int dp^0 dk^0 dq^0 G(k^0,p^0,q^0) 
 \frac{e^{-\ii
 T(2k^0+p^0+q^0-2\mo)}}{(k^0+p^0-\mo-\izero)(k^0+q^0-\mo-\izero)}\\
 \label{ff}
+&\frac{1}{4}\limT\int dp^0 dk^0 dq^0 G(k^0,p^0,q^0) 
\frac{e^{\ii T(p^0-q^0)}}{(k^0+p^0-\mo+\izero)(k^0+q^0-\mo-\izero)}\\
\label{gg}
+&\frac{1}{4}\limT\int dp^0 dk^0 dq^0 G(k^0,p^0,q^0) 
\frac{e^{\ii T(q^0-p^0)}}{(k^0+p^0-\mo-\izero)(k^0+q^0-\mo+\izero)}.
\end{align}
\label{chiii4}%
\end{subequations}

Integrands in terms (\ref{bb})--(\ref{gg}) involve factors 
\be
\frac{e^{\pm \ii Th^0+\cdots}}{\cdots+h^0\pm \izero},
\label{argumenty}
\ee
where $h^0$ variables are timelike components of $4$-momenta appearing in 
expressions for propagators. If we now  integrate each term 
on  semicircular contours in  upper $(+)$ and lower $(-)$ half-planes 
of complex $h^0$, we will see that poles of (\ref{argumenty})  
do not contribute to such contour integrals.
Thus, only 
poles of the $G$ function contribute, but they  turn  exponential
terms into the form similar to (\ref{exp0}). 
For  $M$, $M'$, and $M''$ listed below (\ref{chiii1}), one can then easily argue   that 
 (\ref{bb})--(\ref{gg})  are removed by the limit $\barelimT$.

All in all, we get 
\be
\chi_\Rzymskie{3}
= \pi^2\int dp^0 G(\mo-p^0,p^0,p^0).
\label{chiii}
\ee

\section{Pauli-Villars regularization}
\label{Pauli_sec}
We will discuss here  technicalities related to implementation of  the Pauli-Villars 
regularization through introduction of ghost fields, whose
interaction-picture propagators are \cite{Gupta}
\begin{align}
&
\tilde S(x-y)=
\la\tilde0|\T\tilde\psi_I(x)\overline{\tilde\psi}_I(y)|\tilde0\ra=
\ii\int\frac{\dd{p}}{(2\pi)^4}\frac{\gamma\cdot p+\Lambda}{p^2-\Lambda^2+\izero}e^{-\ii p\cdot(x-y)},
\label{vbnml2}\\
&\tilde D_{\mu\nu}(x-y)=\la\tilde0|\T\tilde A^I_\mu(x)\tilde A^I_\nu(y)|\tilde0\ra
=\ii \int\dddd{p}\frac{e^{-\ii p\cdot (x-y)}}{p^2-\Lambda^2+\izero}
\B{\eta_{\mu\nu}+\xxii\frac{p_\mu p_\nu}{p^2-\Lambda^2/\xi+\izero}}.
\label{yhnnhy}
\end{align}
A quick look at (\ref{prop_fer}) and (\ref{prop_el}) reveals that while 
$S(x-y)$ and $\tilde S(x-y)$ differ only in masses, 
$D_{\mu\nu}(x-y)$ and $\tilde D_{\mu\nu}(x-y)$ differ  in both masses and  overall signs.

Modification of (\ref{fghj}) by (\ref{repl}) asks for evaluation of
\be
\la\bullet|  \T O_I \tilde{\cal H}^I_\IN(x)\tilde{\cal H}^I_\IN(y)|\bullet\ra=
\eo^2{\cal M}_E{\cal M}_D,
\label{MM}
\ee
where  matrix elements involving either real or ghost electromagnetic (Dirac field) 
 operators are denoted as  ${\cal M}_E$   (${\cal M}_D$).
Their indices  are suppressed for the sake of brevity.
Expressions for ${\cal M}_E$ and ${\cal M}_D$ can be easily 
derived  with
the help of Wick's theorem. During their evaluation, one must  keep in mind that 
ghost  fields  follow bosonic statistics. Moreover, it is 
worth to remember that operators $O_I$ are normal ordered (the same
comment  applies to their ghost counterparts $\tilde O_I$ and  to $\tilde{\cal
H}^I_\IN$). Normal ordering of all these operators substantially simplifies  resulting
expressions.

{\bf Dirac field  operators}. Taking  $O=J^i_\sp,J^i_\orb$,  we obtain 
\be
\label{elMfer}
{\cal M}_E= D_{\mu\nu}(x-y)+\tilde D_{\mu\nu}(x-y),
\ee
\ba
{\cal M}_D=&
\la\0s|\T O_I\LN\overline{\psi}_I(x)\gamma^\mu\psi_I(x)\RN\LN\overline{\psi}_I(y)\gamma^\nu\psi_I(y)\RN|\0s\ra\\ 
+&\la\0s|O_I|\0s\ra \trr{\tilde S(y-x)\gamma^\mu\tilde S(x-y)\gamma^\nu}.
\ea
These two formulae also hold when   the unit operator is substituted for $O$. This
observation is useful  during studies  of fermionic spin angular momentum of
the electron,
where the denominator of (\ref{fghj_b}) non-trivially contributes.

{\bf Electromagnetic operators}. For  $O=J^i_\spel, J^i_\orbel,J^i_\xi$, we get  
\begin{align}
\label{elMel}
&{\cal M}_E=\la0|\T O_I A^I_\mu(x) A^I_\nu(y)|0\ra,\\
&{\cal M}_D= {\cal F}^{\mu\nu}(x,y)+\Vol\trr{\tilde S(y-x)\gamma^\mu\tilde S(x-y)\gamma^\nu},
\label{ferMel}
\end{align}
where ${\cal F}^{\mu\nu}$ is given by (\ref{ploik}).
Note that the last term of (\ref{ferMel}) is
$\sz$-independent, and so it has no influence on angular momentum
of the electron due to  reasons explained below (\ref{toot}).

Having these results, one can easily show that replacements (\ref{repl}), when 
performed on (\ref{fghj}), lead to
\begin{align}
\label{ferfer}
&\expval{J^i_\chi}{\Opr}{\lambda} \to \expval{J^i_\chi}{\Opr}{\lambda}-\expval{J^i_\chi}{\Opr}{\Lambda} \ \for \ \chi=\sp,\orb,\\
&\expval{J^i_\chi}{\Opr}{\lambda} \to
\expval{J^i_\chi}{\Opr}{\lambda} \ \for \ \chi=\spel,\orbel,\xi,
\label{elel}
\end{align}
where  the superscript $\lambda$ reminds us that before introduction of  ghost fields 
our calculations have already been IR-regularized.
Thus, while angular momenta listed in  (\ref{ferfer}) are  regularized by modification (\ref{PVL}) 
of the Lagrangian density, the ones from  (\ref{elel}) are not.
We mention in passing that  (\ref{ferfer}) follows from the fact that (\ref{elMfer}) can be written
as $D_{\mu\nu}(x-y)-\ltoL$. 

To fix the problem caused by (\ref{elel}), we consider expectation values of differences of angular
momentum operators and their ghost counterparts. This asks for evaluation of
the analog of (\ref{MM}),
\be
\la\bullet|  \T \tilde O_I \tilde{\cal H}^I_\IN(x)\tilde{\cal H}^I_\IN(y)|\bullet\ra=
\eo^2\tilde{\cal M}_E\tilde{\cal M}_D,
\label{tMM}
\ee
leading to the following set of expressions.

{\bf Ghost Dirac field operators}. Taking $\tilde O=\tilde J^i_\sp,\tilde J^i_\orb$, we obtain
\be
 \tilde{\cal M}_E=D_{\mu\nu}(x-y)+\tilde D_{\mu\nu}(x-y),
\ee
\ba
\tilde{\cal M}_D=
&\frac{\eta^{\mu0}}{(2\pi)^3} \la\tilde 0|\T\tilde  O_I\LN\overline{\tilde\psi}_I(y)\gamma^\nu\tilde\psi_I(y)\RN|\tilde 0\ra \\
+&\frac{\eta^{\nu0}}{(2\pi)^3} \la\tilde 0|\T \tilde O_I\LN\overline{\tilde\psi}_I(x)\gamma^\mu\tilde\psi_I(x)\RN|\tilde 0\ra \\
+&\Vol\la\tilde 0|\T  \tilde O_I\LN\overline{\tilde\psi}_I(x)\gamma^\mu\tilde\psi_I(x)\RN
\LN\overline{\tilde\psi}_I(y)\gamma^\nu\tilde\psi_I(y)\RN|\tilde0\ra,
\label{ghMfer}
\ea
where we have used (\ref{ext_contractions}) and (\ref{ubaru0}) to arrive at (\ref{ghMfer}).

{\bf Ghost electromagnetic operators}. For $\tilde O=\tilde J^i_\spel, \tilde J^i_\orbel,\tilde J^i_\xi$,
we get
\begin{align}
\label{ghelMel}
&\tilde{\cal M}_E=\la\tilde0|\T \tilde O_I \tilde A^I_\mu(x) \tilde
A^I_\nu(y)|\tilde0\ra,\\
&\tilde{\cal M}_D= {\cal F}^{\mu\nu}(x,y)+\Vol\trr{\tilde S(y-x)\gamma^\mu\tilde S(x-y)\gamma^\nu}.
\label{ghferMfer}
\end{align}

Using  (\ref{MM})--(\ref{ferMel}) and  (\ref{tMM})--(\ref{ghferMfer}),  one can show that if we impose on (\ref{fghj}) replacements
\be
J^i_\chi\to J^i_\chi-\tilde J^i_\chi
\label{JJtJ}
\ee
and (\ref{repl}), 
then such modifications will result in
\be
\expval{J^i_\chi}{\Opr}{\lambda} \to \expval{J^i_\chi}{\Opr}{\lambda}-\expval{J^i_\chi}{\Opr}{\Lambda} \
\for \ \chi=\sp,\orb,\spel,\orbel,\xi.
\label{dreg}
\ee
 Two comments are in order now.

First,  ghost operator subtraction (\ref{JJtJ}) does not affect expectation values of  angular momentum 
operators built of Dirac fields,
which are regularized by mere addition of ghost fields to the Lagrangian density,
see (\ref{ferfer}).  The easiest way
to see this is to combine the observation that whole (\ref{ghMfer}) is $\sz$-independent 
with   arguments  presented below (\ref{toot}). 

Second,  ghost operator subtraction  (\ref{JJtJ}) 
leads to regularization of  angular momentum operators composed of electromagnetic operators,
for which (\ref{dreg}) can be understood by noting that  (\ref{ghelMel}) is obtained by
performing the  transformation $\lambda\to\Lambda$  on (\ref{elMel}).

\section{Evaluation of integrals}
\label{Integrals_app}
We evaluate here definite integrals from  (\ref{d1_s}) and (\ref{d2new}). To this aim,
we need the following indefinite integrals
\begin{multline}
4\int ds (1-s)
\B{
\ln\Delta_\chi + \frac{1+s^2}{\Delta_\chi}  }=2s(\tchi^2-4)
-\BB{(\tchi^2-2)^2 + 2(1-s)^2}\ln\BB{(1-s)^2+s\tchi^2}\\+
\frac{2\tchi(\tchi^4-6\tchi^2+12)}{\sqrt{4-\tchi^2}}\arctan\frac{\tchi^2-2(1-s)}{\tchi\sqrt{4-\tchi^2}}+\text{const}
\label{I1}
\end{multline}
and
\begin{multline}
4\int ds\B{
s\ln\Delta_\chi+\frac{2(2-s)(1-s)s}{\Delta_\chi}}=2s(s-3\tchi^2-6)+
(3\tchi^4+2s^2-6)\ln\BB{(1-s)^2+s\tchi^2}\\-
\frac{6\tchi(\tchi^4-2\tchi^2-4)}{\sqrt{4-\tchi^2}}\arctan\frac{\tchi^2-2(1-s)}{\tchi\sqrt{4-\tchi^2}}+\text{const},
\label{I2}
\end{multline}
where 
$\tchi=\chi/\mo$.

These expressions can be used for any  $0<\tchi^2<4$. 
For $\tchi^2>4$, the following replacements 
\begin{align}
&\sqrt{4-\tchi^2}\to\ii\sqrt{\tchi^2-4}, \\
&\arctan\frac{\tchi^2-2(1-s)}{\tchi\sqrt{4-\tchi^2}}\to
-\ii\arctanh\frac{\tchi^2-2(1-s)}{\tchi\sqrt{\tchi^2-4}}
\end{align}
should be employed. They make  right-hand sides of (\ref{I1}) and (\ref{I2}) real.
Analogical replacements   are also meant to be applied  below.

Using (\ref{I1}) and (\ref{I2}), we find 
\begin{subequations}
\begin{align}
& \int_0^1 ds (1-s)
\BB{
\ln\frac{\Delta_\Lambda}{\Delta_\lambda} + (1+s^2) \B{\frac{1}{\Delta_\Lambda} -
\frac{1}{\Delta_\lambda}}}
=I_1(\tilde\lambda)-I_1(\tilde\Lambda)-2\ln\frac{\Lambda}{\lambda},\\
&I_1(x)=\frac{x^2-4}{2}(x^2\ln x-1)-
\frac{x^4-6x^2+12}{2}\frac{x}{\sqrt{4-x^2}}\arctan\frac{\sqrt{4-x^2}}{x}
\end{align}
\label{II1}%
\end{subequations}
and
\begin{subequations}
\begin{align}
&\int_0^1 ds\left[
s\ln\frac{\Delta_\lambda}{\Delta_\Lambda}
+2(2-s)(1-s)s\B{\frac{1}{\Delta_\lambda}
-\frac{1}{\Delta_\Lambda}}\right]=I_2(\tilde\lambda)-I_2(\tilde\Lambda)+2\ln\frac{\Lambda}{\lambda},\\
&I_2(x)=\frac{3x^2}{2}(x^2\ln x-1)-\frac{3(x^4-2x^2-4)}{2}\frac{x}{\sqrt{4-x^2}}\arctan\frac{\sqrt{4-x^2}}{x},
\end{align}
\label{II2}%
\end{subequations}
respectively.


%

\end{document}